%% file: DDF_IT_jan_16.tex
\documentclass[11pt,draftcls,onecolumn]{IEEEtran}
\usepackage{amsmath,amssymb,amsfonts,color,graphicx,eucal}
\usepackage{epsfig}
\usepackage{calc}
\input{macros}

\newcommand{\herm}{{\sf H}}
\newcommand{\transp}{{\sf T}}

\newcommand{\beq}{\begin{equation}}
\newcommand{\eeq}{\end{equation}}

\newcommand{\bea}{\begin{eqnarray}}
\newcommand{\eea}{\end{eqnarray}}
\newcommand{\bean}{\begin{eqnarray*}}
\newcommand{\eean}{\end{eqnarray*}}

\newcommand{\bit}{\begin{itemize}}
\newcommand{\eit}{\end{itemize}}
\newcommand{\ben}{\begin{enumerate}}
\newcommand{\een}{\end{enumerate}}

\newtheorem{thm}{Theorem}

\newtheorem{defn}{Definition}
\newtheorem{lem}[thm]{Lemma}

\def\BibTeX{{\rm B\kern-.05em{\sc i\kern-.025em b}\kern-.08em
    T\kern-.1667em\lower.7ex\hbox{E}\kern-.125emX}}

\begin{document}

\title{Coding and Decoding for the Dynamic Decode and Forward Relay Protocol}

\author{K.~Raj~Kumar and Giuseppe Caire%
\thanks{The authors are with the Department of Electrical Engineering - Systems,
University of Southern California, Los Angeles, CA 90089, USA ({\tt
\{rkkrishn,caire\}@usc.edu}).}%
\thanks{The material in this paper was presented in part at the forty-fifth annual Allerton conference
on Communication, Control, and Computing, Illinois, Sept. 26 - 28,
2007.}%
\thanks{This work was partially supported by
NSF Grant No. CCF-0635326 and by the Oakley fellowship from the
Graduate School at the University of Southern California.}}
\maketitle

\markboth{Submitted to IEEE Trans. Inform. Theory, Jan. 2008}{}

\begin{abstract}
We study the Dynamic Decode and Forward (DDF) protocol for a single
half-duplex relay, single-antenna channel with quasi-static fading.
The DDF protocol is well-known and has been analyzed in terms of the
Diversity-Multiplexing Tradeoff (DMT) in the infinite block length
limit. We characterize the finite block length DMT and give new
explicit code constructions. The finite block length analysis
illuminates a few key aspects that have been neglected in the
previous literature: 1) we show that one dominating cause of
degradation with respect to the infinite block length regime is the
event of decoding error at the relay; 2) we explicitly take into
account the fact that the destination does not generally know a
priori the relay decision time at which the relay switches from
listening to transmit mode. Both the above problems can be tackled
by a careful design of the decoding algorithm. In particular, we
introduce a decision rejection criterion at the relay based on
Forney's decision rule (a variant of the Neyman-Pearson rule), such
that the relay triggers transmission only when its decision is
reliable. Also, we show that a receiver based on the Generalized
Likelihood Ratio Test rule that jointly decodes the relay decision
time and the information message achieves the optimal DMT. Our
results show that no cyclic redundancy check (CRC) for error
detection or additional protocol overhead to communicate the
decision time are needed for DDF. Finally, we investigate the use of
minimum mean squared error generalized decision feedback equalizer
(MMSE-GDFE) lattice decoding at both the relay and the destination,
and show that it provides near optimal performance at moderate
complexity.
\end{abstract}

\section{Introduction} \label{intro.sect}

Employing multiple antennas at the transmitter and the receiver of
wireless communications is known to provide significant benefits in
terms of both throughput (multiplexing gain) and reliability
(diversity gain) (see \cite{ZheTse} and references therein). When
physical constraints limit the number of antennas that can be
installed on a single wireless device (e.g., small sensors in sensor
networks), the usage of cooperative wireless relay protocols is a
promising alternative strategy. In these protocols, two or more
terminals cooperate in order to mimic a super-user with multiple
antennas.

The relay channel was introduced by van der Meulen
\cite{Meu_appl_prob} and was studied in detail by Cover and El Gamal
\cite{CovGam}, who characterized the capacity for the discrete
memoryless as well as for the Gaussian {\em degraded} cases. The
relay channel with fading was examined by Sendonaris et al.,
\cite{SenErkAaz}, where an achievable rate region was provided. In
the case of slow fading, the {\em outage} behavior of half-duplex
wireless relay channels was studied by Laneman et al.,
\cite{LanTseWor}, and simple {\em cooperative diversity} protocols
for signalling across these channels (such as {\em amplify and
forward} and {\em decode and forward}) were introduced. In
\cite{AzaGamSch}, Azarian et al. used the {\em
diversity-multiplexing tradeoff} (DMT) formulation of \cite{ZheTse} to study the
outage behavior of slowly-fading relay channels in the high-SNR regime, and
also introduced new classes of protocols such as the {\em
non-orthogonal amplify and forward} (NAF) and the {\em dynamic
decode and forward} (DDF). An improved DDF protocol based on code
superposition was later proposed in \cite{prasad}.
The DDF protocol for the single relay case was subsequently
studied in \cite{MurAzaGam}, where simplified variants of the protocol
were introduced and some code design issues were addressed.
Code design for the DDF protocol is also addressed in the recent
contribution \cite{EliKum_arxiv}.

The present paper also focuses on the DDF protocol for the
half-duplex, single relay single-antenna case. With respect to
\cite{MurAzaGam} and \cite{EliKum_arxiv}, we analyze explicitly the
achievable DMT of practical codes with finite block length and
propose a simple DMT optimal code construction that makes use of
approximately universal codes for the parallel channel and of the
Alamouti code. Approximately universal codes for the parallel
channel may be obtained either from using a QAM base alphabet and a
suitable unitary precoding matrix (lattice codes) or from
permutation codes derived from universally decodable matrices (UDM)
\cite{TavVis_IT,GanVon}. We treat both cases and give construction
examples and comparisons. Remarkably, our codes perform very close
to the outage probability and have generally lower decoding
complexity than those previously proposed.

Furthermore, we discuss two often neglected issues: 1) the effect of decoding errors at the relay, and how to mitigate it;
2) the fact that the destination does not generally know a priori the relay decision time.
In order to tackle 1), we introduce a decision rejection criterion at the relay, such that the
relay triggers transmission only when its decision is reliable.  We show that the
Forney's decision rule (a variant of Neyman-Pearson rule) yields almost optimal performance
with practical finite length codes, while previously proposed options suffer from significant degradation.
In order to tackle 2), we treat the channel ``seen at destination'' as a compound channel,
where each compound member corresponds to a different relay decision time. We prove
that a receiver based on the Generalized Likelihood Ratio Test (GLRT) rule,
that jointly decodes the relay decision time and the information message,
achieves the optimal DMT. We also show that a simpler scheme that performs separate
detection of the relay decision time, by ignoring the structure of the coded signal and treating it as random,
is generally suboptimal and it becomes optimal only in the limit of infinite block length.
As an aside, our results show that no side information channel
or additional protocol overhead is needed in order to inform the destination about the
relay decision time. This may yield to much simplified actual protocol design for the DDF scheme,
at the cost of an augmented decoder at the destination.

With the lattice codes advocated in this paper, the decoder at the
relay has to solve a closest lattice point problem with a rank
deficient lattice matrix. It is well-known that standard sphere
decoding \cite{VitBou,HasVik} yields exponential complexity in this
case. In order to address this problem (again, often neglected in
the current literature) we advocate the use of the minimum mean
squared error generalized decision feedback equalizer (MMSE-GDFE)
lattice decoder of \cite{MurGamDamCai,DamGamCai_underdet_CISS}. Via
simulation of the performance of our explicitly constructed codes,
we demonstrate that this lattice decoder is able to provide near
optimal performance at moderate complexity.

In Section~\ref{problem.sect}, we introduce the system model we work
with and review relevant previous results. Section~\ref{outage.sect}
presents the main result of the present paper, a characterization of
the DMT of the DDF protocol for finite block length. Explicit code
constructions that achieve this DMT are provided in
Section~\ref{code-construction.sect}, and methods to enable error
detection at the relay and low complexity decoding of these codes
are also dealt with.

\section{Problem definition and background} \label{problem.sect}

\subsection{System model} \label{model.sect}

\begin{figure}[h]
\begin{center}
\includegraphics[width=8cm]{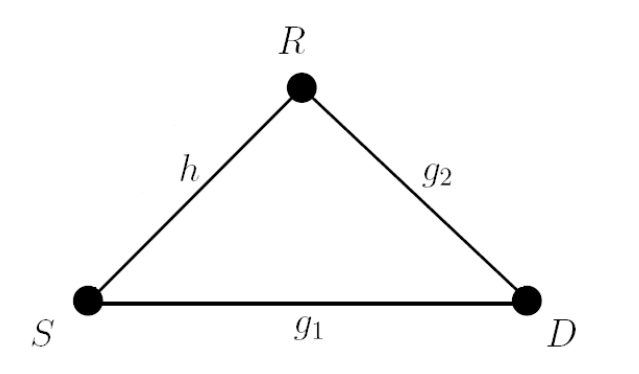}
\caption{The single-antenna single relay fading channel. \label{fig:single_relay}}
\end{center}
\end{figure}

We consider the single relay channel shown
in Fig.~\ref{fig:single_relay}, where S, R and D denote the source,
relay and destination, and $h, \ g_1$ and $g_2$ denote the fading coefficients
between the source-relay, source-destination and relay-destination terminals, respectively.

The channel fading coefficients are i.i.d. $\mathcal{CN}(0,1)$
random variables, corresponding to i.i.d. Rayleigh fading. Following
the standard outage setting \cite{LanTseWor,AzaGamSch,ZheTse}, we
assume that the channel coherence time is considerably larger than
the allowed decoding delay. Invoking a time-scale decomposition
argument (see for example \cite{TseVis}) this setting is modeled by
the so-called quasi-static fading channel, where the channel
coefficients are random but remain constant over the whole duration
of a codeword, although the latter can be very large. We consider
slotted transmission where a source codeword spans $M$ slots of
length $T$ symbols each, resulting in a total block length of $MT$.

The relay operates in half-duplex mode. In decode and forward
protocols, the block of length $MT$ symbols is split into two
phases. In the first phase the relay is in listening mode and
receives the signal from the source. At a certain instant, referred
to as the {\em decision time} in the following, the relay tries to
decode the source information message. In the second phase, from the
decision time to the end of the block, the relay switches to
transmit mode and sends symbols to help the destination decode the
source message. The DDF protocol is characterized by the fact that
the decision time is not fixed a priori. On the contrary, the relay
decides when to decode and switch to transmit mode depending on the
channel coefficient $h$ and the received signal. Therefore,  the
decision time is a random variable $\Mc$. Without loss of
generality, we restrict the decision time to coincide with the end
of a slot\footnote{Notice that $T$ is a design parameter. Letting $T
= 1$ provides an unrestricted decision time. In this way, there is
no loss of generality in this assumption.}, i.e., $\Mc$ takes on
values in the set $\{1,2,\ldots,M\}$, where $\Mc = M$ corresponds to
the case where the relay does not help the destination. During phase
1 (listening phase) the signal received by the relay is
\begin{equation} \label{eq:recd_relay}
y_{r,k} = h x_{s,k} + v_k, \ k = 1,2,\hdots,\Mc T,
\end{equation}
and the signal received by the destination is
\begin{equation} \label{eq:recd_destination1}
y_k = g_1 x_{s,k}  + w_k, \ k = 1,2,\hdots,\Mc T.
\end{equation}
During phase 2 (relay transmit phase), the signal received by the destination is
\begin{equation} \label{eq:recd_destination2}
y_k = g_1 x_{s,k} + g_2 x_{r,k} + w_k, \ k = \Mc T+1, \Mc T+
2,\hdots,MT.
\end{equation}
Here, $\xv_s = [x_{s,1} \cdots x_{s,MT}]^\transp$ denotes the source
codeword, drawn from a code $\mathcal{X}_s \subset \CC^{MT}$ of rate
$R$ bits per symbol. Without loss of generality, we may assume that
the symbols $x_{r,k}$ transmitted by the relay are from an auxiliary
code $\mathcal{X}_r \subset \CC^{MT}$ with rate $R$ and block length
$MT$, but only the last $(M-\Mc)T$ symbols of a codeword are
effectively transmitted in phase 2, while in phase 1 the relay
transmitter is idle because of the half-duplex constraint.

The noise at the relay and destination, denoted by $v_k \sim
\Cc\Nc(0,\sigma_v^2)$ and $w_k \sim \Cc\Nc(0,\sigma_w^2)$, form two
white mutually independent sequences. We impose the same per-symbol
average power constraint for both the source and the relay, given by
\[ \mathbb{E} \left[ |x_{s,k}|^2 \right], \ \mathbb{E} \left[ |x_{r,k}|^2 \right] \leq E, \]
where $E$ denotes the symbol energy, and define the SNRs of the S-D and
the S-R links to be $\rho = E/\sigma_w^2$ and $\rho' = E/\sigma_v^2$, respectively.

For later use, we introduce the following notation: let $\yv_{i}^j$,
$\yv_{r,i}^j$, $\xv_{s,i}^j$ and $\xv_{r,i}^j$, each $\in
\CC^{(j-i)T}$, denote respectively the received signals at the
destination and at the relay from symbol time $iT+1$ to $jT$, the
source transmit signal from time $iT+1$ to $jT$ and the relay
transmit signal from time $iT+1$ to $jT$, where the latter is
assumed to be zero for all times $k \leq \Mc T$. The quantities
$\wv_i^j$ and $\vv_i^j$ are defined similarly.

\subsection{Diversity-Multiplexing Tradeoff} \label{dmt.sect}

A compact and convenient characterization of the tradeoff between
rate and reliability of quasi-static fading channels in the high-SNR regime
is provided by the DMT introduced in \cite{ZheTse}.
In this framework, rate and reliability are
quantified in terms of the {\em diversity gain} $d$ and {\em spatial
multiplexing gain} $r$. A family of coding systems, each of which
operates at SNR $\rho$ with rate $R(\rho)$ and error probability
$P_e(\rho)$, achieves a point $(r,d)$ on the DMT plane if
\[ \lim\limits_{\rho \rightarrow \infty} \frac{R(\rho)}{\log
\rho}= r, \ \lim\limits_{\rho \rightarrow \infty} \frac{\log
P_e(\rho)}{\log \rho}= -d.\] This latter relation is written as
$P_e(\rho) \doteq \rho^{-d}$ in the exponential equality notation of
\cite{ZheTse}.

We will use the DMT as our performance metric when we analyze cooperative diversity protocols.
It is clear that the DMT of the MIMO channel with one receive and two
transmit antennas provides an upper bound to the performance of any
relay protocol for the channel of Fig.~\ref{fig:single_relay}.
This bound, known as the transmit diversity bound \cite{LanTseWor}, is given by
\[ d_{\text{tx.div.bd.}}(r) = 2(1-r).\]
The DMT of the DDF protocol, proposed and analyzed in \cite{AzaGamSch},
is given by
\begin{equation} \label{eq:DMT_DDF}
d^*(r) = \left\{
\begin{array}{cc}
2(1-r), & 0 \leq r \leq \frac{1}{2}\\
(1-r)/r, & \frac{1}{2} \leq r \leq 1
\end{array}.
\right.
\end{equation}
This result is obtained by analyzing the information outage probability with Gaussian inputs, and it is
achievable (e.g., by using a Gaussian random coding argument) in the limit of both
$M \rightarrow \infty$ and $T \rightarrow \infty$.
The relay decision time is given by
\begin{equation} \label{ddf-classic}
\Mc = \min \left\{ M, \frac{MR}{\log (1 + |h|^2 \rho')} \right\},
\end{equation}
i.e., $\Mc$ is set to the minimum $m = 1,2,\ldots,M - 1$ such that
the mutual information between $\xv_{s,0}^{m}$ and $\yv_{r,0}^{m}$
for fixed and known $h$, given by $mT \log (1 + |h|^2 \rho')$,
exceeds the number of information bits per message $MTR$.
If such an $m$ exists, the relay triggers the decoding of the whole
information message and switches to the transmission mode. If no
such $m$ exists, then $\Mc = M$ and the relay remains silent. Both
the limit of large $M$ and $T$ are necessary to achieve the DDF DMT
in (\ref{eq:DMT_DDF}). In fact, the normalized decision time $\Mc/M$
must converge to a continuous random variable distributed in $[0,1]$
and, for every decision time $\Mc = m$, the number of symbols $mT$
received by the relay must be arbitrarily large, such that the
decoding error event coincides with the information outage error
event. In this way, the corresponding probability of decoding error
is arbitrarily close to the information outage probability
\[ P \left (\log(1 + |h|^2 \rho') \leq \frac{MR}{m} \right ), \]
and the probability of undetected error (i.e., the relay accepts a wrong decision) is arbitrarily small.
In brief, $T \rightarrow \infty$ is necessary in order to fix the optimal decision time based only on the channel
strength $|h|^2$ and be sure (with arbitrarily high probability) that the decoded message is the correct one.

We should also notice that, in the limit of $T \rightarrow \infty$,
the outage probability does not depend on the knowledge of $h$ at
the relay decoder and of $(g_1,g_2)$ at the destination decoder (see
for example \cite{BigProSha}). On the other hand, a common
assumption made in previous works is that the destination knows
exactly the relay decision time $\Mc$. In practice, this assumption
requires some form of protocol to provide side information to the
destination. In the DMT analysis, one should pay great care to
ensure that the error probability of such side information protocol
does not dominate the decoding error probability, i.e., in designing
any side information protocol we must ensure that its probability of
error decreases not slower than $\rho^{-d^*(r)}$.

Practical code design for the DDF protocol considers finite,
possibly very short,  $M$ and $T$. In the following, we will make an
explicit assumption of perfect receiver channel state information
(CSIR), that is relatively easy to acquire using pilot symbols and
is a common assumption in the DMT analysis of even finite-length
codes (see \cite{ZheTse} and \cite{TseVis}). On the contrary, we
explicitly address the fact that the destination does not know a
priori the relay decision time $\Mc$ and tackle this problem by
analyzing an augmented decoder based on the GLRT rule.

\subsection{Existing DDF code designs} \label{previous.sect}

In \cite{MurAzaGam}, a variant of the DDF protocol is proposed where
the relay code $\mathcal{X}_r$ is such that the signal received at
the destination reduces to an Alamouti constellation \cite{Ala}. We
will refer to this scheme as the ``Alamouti-DDF'' scheme, and review
it briefly in the sequel since we make use of the same approach.
With the Alamouti-DDF, assuming that the relay decodes correctly at
the decision time $\mathcal{M} = m$, the signal transmitted by the
relay at time $k$ is given by~\cite{MurAzaGam}
\begin{equation} \label{eq:Ala_scheme}
x_{r,k} = \left\{
\begin{array}{cc}
x^*_{s,k+1}, & k = mT+1, mT+3, \hdots\\
-x^*_{s,k-1}, & k = mT+2, mT+4, \hdots
\end{array}, \right.
\end{equation}
which reduces the signal seen by the destination for $mT+1 \leq k
\leq MT$ to an Alamouti constellation. Through linear processing of
the received signal $\yv_0^{M}$, the destination obtains the
sufficient statistics for decoding, given by
\begin{equation} \label{eq:AlaDDF_channel}
\tilde{y}_k =
\left\{
\begin{array}{cc}
g_1 x_{s,k} + w_k, & k = 1, \hdots, mT\\
\sqrt{|g_1|^2 + |g_2|^2} x_{s,k} + \tilde{w}_k, & k = mT+1, \hdots, MT
\end{array} \right. ,
\end{equation}
where the statistics of $\tilde{w}_k$ are identical to those of
$w_k$. In this case, it is easy to see that the mutual information
per symbol at the destination, for $\Mc = m$ and i.i.d. Gaussian
inputs, is given by
\begin{equation} \label{abuse1}
\frac{m}{M} \log \left (1 + |g_1|^2 \rho \right ) + \frac{M - m}{M} \log \left (1 + (|g_1|^2 + |g_2|^2) \rho \right )
\end{equation}
and coincides with that of the original DDF scheme defined by
(\ref{eq:recd_destination1}) and (\ref{eq:recd_destination2}), when
the codebooks $\Xc_s$ and $\Xc_r$ are also drawn independently from
an i.i.d. Gaussian ensemble. Hence, the Alamouti-DDF modification
entails no loss in DMT compared to the original DDF protocol
\cite{MurAzaGam}.

%

\section{DMT of the DDF Protocol with finite length} \label{outage.sect}

In this section, we characterize the achievable DMT of the DDF
protocol with finite $M$ and $T$. First, we find an upper bound on
the DMT by letting $T \rightarrow \infty$, assuming that the
destination has perfect knowledge of the relay decision time $\Mc$,
and using outage probability. Then, we shall analyze the performance
of Gaussian random codes with finite length, with the assumption
that the destination has no knowledge of $\Mc$, and find a lower
bound that matches the upper bound.

Since for i.i.d. Gaussian inputs the Alamouti-DDF yields the same
mutual information as DDF, as far as outage probability is concerned
we can refer to the channel defined in \eqref{eq:AlaDDF_channel}.
This is a set of parallel channels for $m = 1,\ldots,M$, with
dependent channel gains. In particular, there are two types of
sub-channels: one representing the S-D link, and another set
representing the composite (S,R)-D link (except for the case when $m
= M$, which corresponds to when the relay remains inactive for the
whole block; in this case, only the S-D link appears). The switching
point between the two channels is controlled by the random variable
$\mathcal{M}$. We will refer to this channel as a {\em random switch
channel} (RSC). Given a particular switching instant $\mathcal{M} =
m$, we will call the ensuing channel as a {\em $m$-switch channel}
($m$-SC). The RSC belongs to the class of  ``mixed channels'' (see
\cite{Han}), that is, a compound channel with an a priori
probability distribution on the compound members. In this case,
the probability distribution on the channel members
(the $m$-SCs in (\ref{eq:AlaDDF_channel})) is induced by the triple $(\Mc, g_1,g_2)$.

\subsection{Outage probability analysis} \label{dmt-analysis.sect}

We compute the DMT of the RSC defined above for arbitrarily large $T$ under the assumption that
the destination receiver has perfect knowledge of $\Mc$, and hence find an upper bound
on the DMT exponent $d^*_M(r)$ for the finite-length DDF protocol.
This is established by the following theorem.

\vspace{12pt}

\begin{thm} \label{th:DMT_DDF}
The DMT of the single relay DDF scheme with decision times $m =
1,2,\ldots,M$ and finite slot length $T \geq 1$ is upper bounded by
\[ d^*_M(r) \leq d_{\rm out}(r) = \min \limits_{1 \leq m \leq M} \left\{ \overline{d}_m(r) + d_m(r) \right\}, \]
where
\begin{equation} \label{eq:d_bar}
\overline{d}_m(r) = \left\{
\begin{array}{cc}
1 - \frac{Mr}{m-1}, & 0 \leq r \leq \frac{m-1}{M}\\
0, & \frac{m-1}{M} < r \leq \frac{m}{M}\\
\infty, & \frac{m}{M} < r \leq 1
\end{array}
\right. ,
\end{equation}
\begin{equation} \label{dm1}
d_m(r) = \left\{
\begin{array}{ll}
2-2r, & m < \frac{M}{2}\\
\frac{M(1-r)}{m}, & m \geq \frac{M}{2}
\end{array}
\right.
\end{equation}
for $r \geq \frac{1}{2}$, and
\begin{equation} \label{dm2}
d_m(r) = \left\{
\begin{array}{ll}
2-2r, & m < \frac{M}{2}\\
2 - \frac{rM}{M-m}, & \frac{M}{2} \leq m < M(1-r)\\
\frac{M(1-r)}{m}, & m \geq M(1-r)
\end{array}
\right.
\end{equation}
for $r < \frac{1}{2}$.
\end{thm}

\vspace{12pt}

\begin{proof}
Let $\Mc$ denote the random decision time as defined in (\ref{ddf-classic}) and
$P_{out}(r)$ denote the outage probability of the corrsponding RSC.
Also, let $P_{out}^{m-\text{SC}}(r)$ denote the outage probability of the
$m$-SC for given $m$. Then, the law of total probability yields
\begin{equation} \label{eq:outage_DDF}
P_{out} (r) = \sum_{m=1}^M P ( \mathcal{M} = m ) P_{out}^{m-\text{SC}}(r).
\end{equation}
Since in the regime of very high SNR that characterizes the DMT, scaling SNR by a constant does not
change the DMT, we allow both $\rho, \rho' \rightarrow \infty$ and the DMT shall not depend on
the (constant) ratio $\rho'/\rho = \sigma_w^2 / \sigma_v^2$. Define
\begin{eqnarray*}
P_{out} (r) &\doteq& \rho^{-d_{\rm out}(r)},\\
P_{out}^{m-\text{SC}}(r) &\doteq& \rho^{-d_m(r)}, \ 1 \leq m \leq
M,\\
P ( \mathcal{M} = m ) &\doteq& \rho^{-\overline{d}_m(r)}, \ 1 \leq m \leq M.
\end{eqnarray*}
Then, it is clear from (\ref{eq:outage_DDF}) that
\[ d_{\rm out}(r) = \min \limits_{1 \leq m \leq M} \left\{ \overline{d}_m(r) + d_m(r) \right\}. \]
Furthermore, from standard arguments based on Fano inequality \cite{ZheTse} and because
here we are assuming that the destination receiver is enhanced by the side information on
$\Mc$, it is also immediate to conclude that $d_M^*(r) \leq d_{\rm out}(r)$.

It remains to prove (\ref{eq:d_bar}) and \eqref{dm1}, \eqref{dm2}.
Notice that $\overline{d}_m(r)$ is solely a function of the S-R link
and $d_m(r)$ is a function of the R-D and S-D links. We analyze
these quantities separately as follows.

\subsubsection{Analysis of $\rho^{-\overline{d}_m(r)}$}
Let's consider first the case $m < M$. Set $R = r \log \rho$. The probability that the relay decodes after $m$
sub-blocks $P( \mathcal{M} = m ) $, $1 \leq m \leq M - 1$, corresponds
to the event
\[\left\{ mT \log (1 + |h|^2 \rho') > MRT > (m-1)T \log (1 + |h|^2
\rho') \right\}\]
\begin{eqnarray} \label{m-event}
&\Leftrightarrow \left\{ \frac{Mr}{m} \log \rho < \log (1 + |h|^2
\rho') <
\frac{Mr}{m-1} \log \rho \right\}& \nonumber \\
&\Leftrightarrow \left\{ \frac{\rho^{\frac{Mr}{m}}-1}{\rho'} < |h|^2
< \frac{\rho^{\frac{Mr}{m-1}}-1}{\rho'} \right\}.&
\end{eqnarray}

Since $|h|^2$ is exponentially distributed and $\rho' \doteq \rho$,
we compute
\begin{eqnarray*}
P ( \mathcal{M} = m ) &\doteq&
\int_{\rho^{\frac{Mr}{m}-1}}^{\rho^{\frac{Mr}{m-1}-1}} e^{-z} dz\\
&=& e^{-\rho^{\frac{Mr}{m}-1}} - e^{-\rho^{\frac{Mr}{m-1}-1}}.
\end{eqnarray*}
According to the value of the multiplexing gain, we analyze the
above quantity for each $1 \leq m < M$ as follows.
\begin{itemize}
\item $r > \frac{m}{M}$:\\
This corresponds to $\frac{Mr}{m}-1,\frac{Mr}{m-1}-1 > 0$. In this
case\footnote{The notation $P \; \doteq \; \rho^{-\infty}$ indicates that $P$ decreases faster than any polynomial function of $\rho$.}
\[ P ( \mathcal{M} = m ) \doteq \rho^{-\infty}.\]

\item $\frac{m-1}{M} < r \leq \frac{m}{M}$:\\
This corresponds to $\frac{Mr}{m}-1 \leq 0, \ \frac{Mr}{m-1}-1 > 0$.
In this case,
\[ P ( \mathcal{M} = m ) \doteq \rho^{0}. \]

\item $r \leq \frac{m-1}{M}$:\\
This corresponds to $\frac{Mr}{m}-1 \leq 0, \ \frac{Mr}{m-1}-1 \leq
0$. In this case, using a power series expansion, \begin{eqnarray*}
P ( \mathcal{M} = m ) &=& \left[ 1 - \rho^{\frac{Mr}{m}-1} +
\frac{\rho^{2\left(\frac{Mr}{m}-1 \right)}}{2!} + \cdots \right] - \\
& & \left[ 1 - \rho^{\frac{Mr}{m-1}-1} +
\frac{\rho^{2\left(\frac{Mr}{m-1}-1 \right)}}{2!} + \cdots
\right]\\
&\doteq& \rho^{\frac{Mr}{m-1} - 1}.
\end{eqnarray*}
\end{itemize}
A similar analysis for $P(\mathcal{M} = M)$ results in
\[ P\{\mathcal{M}  = M \} \doteq \left\{
\begin{array}{cc}
\rho^{\frac{Mr}{M-1}-1}, \ & 0 \leq r \leq \frac{M-1}{M}\\
\rho^{0}, \ & \frac{M-1}{M} < r \leq 1
\end{array}
\right. .\]
Therefore, the result for all $1 \leq m \leq M$ can be compactly expressed by (\ref{eq:d_bar}),
shown in Fig.~\ref{fig:d_bar}.
\begin{figure}[h]
\begin{center}
\includegraphics[width=10cm]{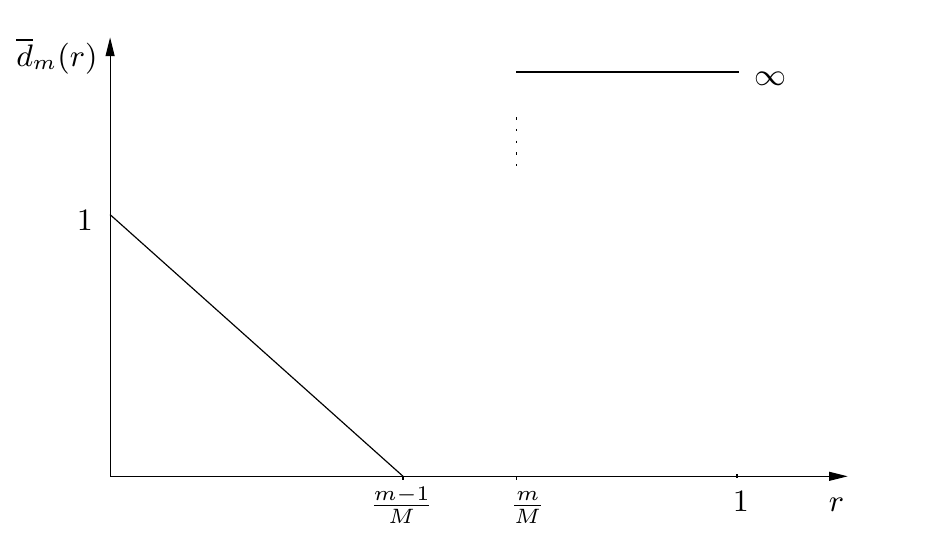}
\caption{Negative $\rho$-exponent of the probability of the relay
decoding after exactly $m$-subblocks. \label{fig:d_bar}}
\end{center}
\end{figure}

\subsubsection{Analysis of $d_m(r)$}
From \eqref{abuse1}, the outage probability of the $m$-SC is given
by
\[ P_{out}^{m-\text{SC}}(r)  = P \left ( \mathcal{I}^{m-\text{SC}} \leq MTR \right ), \]
where $\mathcal{I}^{m-\text{SC}} = mT \log (1 + |g_1|^2 \rho) + (M
- m)T \log [1 + (|g_1|^2 + |g_2|^2)\rho]$.
Defining $|g_1|^2 = \rho^{-\alpha_1}$ and
$|g_2|^2 = \rho^{-\alpha_2}$ and applying standard approximations in the regime of large $\rho$, we eventually obtain
\[
P_{out}^{m-\text{SC}}(r) \; \doteq \;  P \left ( (M-m) \max\{
[1-\alpha_1]_+,[1-\alpha_2]_+ \}  + m [1-\alpha_1]_+
\leq rM \right ),
\]
where $[x]_+ \triangleq \max \{ 0, x \}$.
Since $|g_1|^2$ and $|g_2|^2$ are independent exponential random
variables, the joint pdf of $(\alpha_1,\alpha_2)$ is given by
\[ f(\alpha_1,\alpha_2) \doteq e^{-\rho^{-\alpha_1}-\rho^{-\alpha_2}} \rho^{-\alpha_1-\alpha_2}.\]
Therefore,
\[ P_{out}^{m-\text{SC}}(r) \doteq \int_{\mathcal{B}} \rho^{-\alpha_1 - \alpha_2} d \alpha_1 \ d \alpha_2, \]
where $\mathcal{B}$ is the two-dimensional region defined by the
inequalities $(M-m) \max\{ [1-\alpha_1]_+,[1-\alpha_2]_+ \} +
m [1-\alpha_1]_+ \leq rM$ and $\alpha_i \geq 0 \ \forall \ i$.

Using Varadhan's lemma \cite{DemZei}, we obtain
\begin{equation} \label{eq:infimum}
d_m(r) = \inf\limits_{\mathcal{B}} \left\{ \alpha_1 + \alpha_2
\right\}.
\end{equation}
Define $\beta = \frac{m}{M}$. The region $\mathcal{B}$ is
equivalently defined by
\[ (1-\beta)
\max\{ [1-\alpha_1]_+,[1-\alpha_2]_+ \} + \beta [1-\alpha_1]_+ \leq
r,\]
\[ \alpha_i \geq 0 \ \forall \ i.\]
\begin{figure}[h]
\begin{center}
\includegraphics[width=8cm]{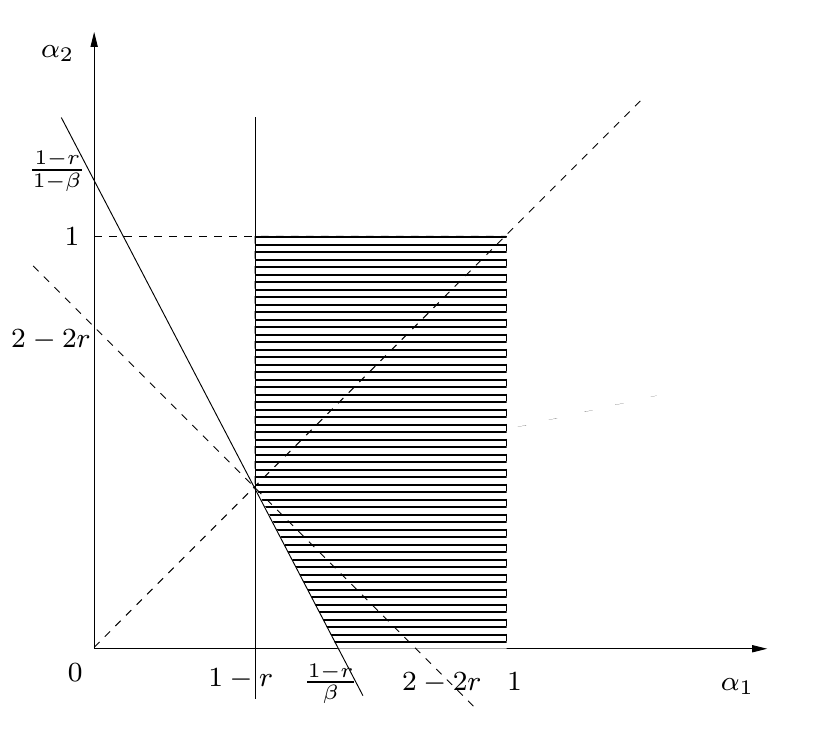}
\caption{The region $\mathcal{B}$. \label{fig:inf_region}}
\end{center}
\end{figure}
It is obvious that we may restrict attention to $\alpha_i \leq 1 \
\forall \ i$ insofar as computing the infimum in \eqref{eq:infimum}
is concerned. We analyze $\mathcal{B}$ according to the following
cases:
\begin{itemize}
\item $\alpha_1 \geq \alpha_2$:\\
We have
\begin{eqnarray*}
&\beta (1 - \alpha_1) + (1-\beta)(1-\alpha_2) \leq r&\\
&\Leftrightarrow \beta \alpha_1 + (1 - \beta) \alpha_2 \geq 1-r.&
\end{eqnarray*}
This line has intercepts $\frac{1-r}{\beta}$ and $\frac{1-r}{1 -
\beta}$ on the $\alpha_1$ and $\alpha_2$ axes respectively.
\item $\alpha_1 < \alpha_2$:\\
We have
\begin{eqnarray*}
&\beta (1 - \alpha_1) + (1-\beta)(1-\alpha_1) \leq r&\\
&\Leftrightarrow \alpha_1 \geq 1-r.&
\end{eqnarray*}
\end{itemize}
The region $\mathcal{B}$ is depicted in Fig.~\ref{fig:inf_region}.
The solution to the problem in \eqref{fig:inf_region} corresponds to
choosing the least non-negative $k$ such that the line $\alpha_1 +
\alpha_2 = k$ touches $\mathcal{B}$. The analysis should be done
according to whether $\frac{1-r}{\beta} \gtrless 2-2r
\Leftrightarrow \beta \lessgtr \frac{1}{2}$ and whether $2-2r
\gtrless 1 \Leftrightarrow r \lessgtr \frac{1}{2}$. It is immediate
from Fig.~\ref{fig:inf_region} that the solution to
\eqref{eq:infimum} when $r \geq \frac{1}{2}$ is at
$(\alpha_1^*,\alpha_2^*) = \left( \frac{1-r}{\beta} ,0 \right)$ for
$\beta \geq 0.5$, and $(\alpha_1^*,\alpha_2^*) = (1-r,1-r)$ for
$\beta < 0.5$. For the case when $r < \frac{1}{2}$, the solution to
\eqref{eq:infimum} is at $(\alpha_1^*,\alpha_2^*) = (1-r,1-r)$ for
$\beta < 0.5$, at $(\alpha_1^*,\alpha_2^*) = \left(1, 1 -
\frac{r}{1-\beta}\right)$ for $\beta \geq 0.5$ and
$\frac{1-r}{\beta} > 1$, and at $(\alpha_1^*,\alpha_2^*) = \left(
\frac{1-r}{\beta} ,0 \right)$ for $\beta \geq 0.5$ and
$\frac{1-r}{\beta} \leq 1$. The final solution is compactly
expressed by \eqref{dm1}, \eqref{dm2}.

This concludes the proof of Theorem \ref{th:DMT_DDF}.
\end{proof}

\subsection{Achievability}

We consider finite length $T$ and no a priori knowledge of $\Mc$ at the destination decoder.
We have the following result:

\vspace{12pt}

\begin{thm} \label{achievability-thm}
The upper bound of Theorem \ref{th:DMT_DDF} is achievable. Therefore,
$d^*_M(r) = d_{\rm out}(r)$.
\end{thm}

\vspace{12pt}

\begin{proof}
We consider the original DDF protocol (not the Alamouti variant) defined by
(\ref{eq:recd_relay}), (\ref{eq:recd_destination1}) and (\ref{eq:recd_destination2}). For this channel we construct
a particular coding scheme and analyze its performance.

{\em Codebook generation:} For given $M$, $T$ and $R$, we generate
$\Xc_s \subset \CC^{MT}$ and $\Xc_r \subset \CC^{MT}$ of cardinality
$\rho^{rMT}$ independently, with i.i.d. components $\sim
\Cc\Nc(0,E)$. We let  $\xv_s(\omega)$  and $\xv_r(\omega)$ denote
the codewords in $\Xc_s$ and in $\Xc_r$, respectively, corresponding
to the information message $\omega \in \{1,\ldots,\rho^{rMT}\}$.

{\em Relay decoding:} We define the relay outage event at slot $m$
as
\begin{equation} \label{relay-outage}
\Oc_m = \left \{ h \in \CC : |h|^2 \leq \frac{\rho^{\frac{rM}{m}} - 1}{\rho'} \right \}
\end{equation}
Differently from the case of arbitrarily large $T$, the relay may
decode in error at time $m$ even though $h \notin \Oc_m$. In the
presence of such {\em undetected error} the relay would switch to
transmit mode and send a codeword corresponding to an incorrect
information message, thus jamming the destination receiver. In order
to avoid this event we consider a bounded distance relay decoding
decision function $\psi_\delta$ defined as follows (see
\cite{GamCaiDam_ARQ}): for $m = 1,\ldots,M-1$, define the regions
$\Sc_m(\omega)$ of all points $\yv \in \CC^{mT}$ for which $\omega$
is the {\em unique} message that is contained in a sphere of squared
radius $mT(1 + \delta)\sigma_v^2$ centered at $\yv$, i.e., $|\yv - h
\xv_{s,0}^{m}(\omega) |^2 \leq mT(1 + \delta)\sigma_v^2$. Then, let
$\psi_\delta(\yv_{r,0}^{m},h) = \widehat{\omega} \in
\{1,\ldots,\rho^{rMT}\}$ if both the following conditions are
satisfied:
\begin{enumerate}
\item $h \notin \Oc_m$;
\item $\yv_{r,0}^{m} \in \Sc_m(\widehat{\omega})$;
\end{enumerate}
(the relay has perfect knowledge of its own channel coefficient $h$,
by the perfect CSIR assumption). If these conditions are satisfied,
then $\Mc = m$ and the relay switches to transmit mode, sending the
signal $\xv_{r,m}^{M}(\widehat{\omega})$ for the remaining part of
the block. Otherwise, it refrains from making a decision and waits
for the next slot.

It should be noticed that the condition 2) above is a test on the typicality of the estimated channel noise.
In fact, if $\omega$ is the transmitted message, we have that
\[ |\yv_{r,0}^{m} - h \xv_{s,0}^{m}(\omega)|^2 = |\vv_0^{m}|^2 \]
is a central chi-squared random variable with $2mT$ degrees of
freedom and mean $mT\sigma_v^2$, that provides an empirical estimate
of the noise variance.

{\em Destination decoding:} The destination is not aware of the
relay decision time $\Mc$. Hence, it makes use of an augmented
decoder that {\em simultaneously} detects the decision time and the
information message according to the GLRT rule:
\begin{equation} \label{GLRT}
\{ \widehat{\omega}, \widehat{m} \} = \arg\max\limits_{\omega,m}
p\left (\yv_0^{M} | \omega ,m,g_1,g_2 \right ).
\end{equation}
where $p(\yv_0^{M}| \omega, m,g_1,g_2)$ is the decoder {\em
likelihood function}, i.e., the pdf of the signal received by the
destination  over the whole block length, under the hypothesis that
the source transmitted the information message $\omega$, that the
relay decision time is  $m$, and given the channel coefficients
$g_1,g_2$ (recall that we assume perfect CSIR).

{\em Error probability analysis:} Let $\Ec$ denote the decoding
error event at the destination and $\Ec_r$ denote the decoding error
event at the relay.\footnote{The complement of an event $\Ac$ is
denoted by $\overline{\Ac}$.} We can write
\begin{eqnarray} \label{cond-error}
P(\Ec) & = & \sum_{m=1}^M P (\Mc = m ) P (\Ec |\Mc = m) \nonumber \\
& = & \sum_{m=1}^M P (\Mc = m ) \left ( P (\Ec, \Ec_r |\Mc = m) +  P (\Ec, \overline{\Ec}_r |\Mc = m) \right ) \nonumber \\
& \leq & \sum_{m=1}^M P (\Mc = m ) \left ( P (\Ec_r |\Mc = m) +   P (\Ec |\overline{\Ec}_r,\Mc = m) P(\overline{\Ec}_r | \Mc = m) \right ) \nonumber \\
& \leq & \sum_{m=1}^M P (\Mc = m ) \left ( P (\Ec_r |\Mc = m) +   P (\Ec |\overline{\Ec}_r,\Mc = m)  \right ).
\end{eqnarray}
First, we bound the effect of the undetected decision error at the
relay. Our analysis follows closely the analysis of the MIMO-ARQ
scheme in \cite{GamCaiDam_ARQ}. In fact, the relay applies a scheme
very similar to ARQ: when it is sure about its decision it stops
receiving and starts transmitting, while if it is not sure about its
decision it waits for the next slot. We have
\begin{eqnarray} \label{relay-undetected-error}
P(\Ec_r | \Mc = m) & = & \rho^{-rMT}
\sum_{\omega=1}^{\rho^{rMT}}
P \left ( \left . \bigcup_{\widehat{\omega} \neq \omega} \left \{ \yv_{r,0}^{m} \in \Sc_m(\widehat{\omega}) \right \} \right | \omega \right ) \nonumber \\
& \leq & P \left ( |\vv_0^{m} |^2 > mT (1 + \delta) \sigma_v^2 \right ) \nonumber \\
& \leq & (1 + \delta)^{mT} e^{-mT\delta},
\end{eqnarray}
where the last line follows from the Chernoff bound on the tail of the chi-squared distribution.
Letting $\delta = \mu \log \rho$, we find
\[ P(\Ec_r | \Mc = m) \; \dot\leq \; \rho^{-mT\mu}. \]
Notice that  $P (\Ec |\overline{\Ec}_r,\Mc = m) \; \dot\geq \;
\rho^{-d_m(r)}$ where $d_m(r)$ is the exponent of the information
outage probability of the $m$-SC channel given in \eqref{dm1},
\eqref{dm2} and is not larger than 2. Hence, it is sufficient to
choose $\mu T > 2$ in order to make the terms  $P(\Ec_r | \Mc = m)$
exponentially irrelevant in (\ref{cond-error}).

Next, let us examine the probabilities $P(\Mc = m)$. Let $\Uc_m =
\bigcup_{\omega=1}^{\rho^{rMT}} \Sc_m(\omega)$ denote the subset of
the relay channel output space $\CC^{mT}$ such that if
$\yv_{r,0}^{m} \in \Uc_m$ then there exists a unique codeword within
the bounded distance decoder's decoding sphere centered at
$\yv_{r,0}^{m}$. For $m = 1$, we have
\begin{eqnarray} \label{cavoli1}
P(\Mc = 1) & = & P \left ( \{ h \notin \Oc_1 \} , \{ \yv_{r,0}^{1} \in \Uc_1\} \right ) \nonumber \\
& \leq & P ( h \notin \Oc_1 ) \nonumber \\
& \doteq & \rho^{-\overline{d}_1(r)}.
\end{eqnarray}
For brevity we let $\Dc_m = \{h \notin \Oc_m\} \cap \{\yv_{r,0}^{m}
\in \Uc_m\}$. Then, for $1 < m < M$, we have
\begin{eqnarray} \label{cavolim}
P(\Mc = m) & = & P \left ( \overline{\Dc}_1, \ldots, \overline{\Dc}_{m-1}, \Dc_m \right ) \nonumber \\
& \leq & P \left ( \overline{\Dc}_{m-1}, \Dc_m \right ).
\end{eqnarray}
For $1 < m < M$, from (\ref{cavolim}) we can write
\begin{eqnarray} \label{cavolim1}
P(\Mc = m) & \leq & P \left ( \left \{ \{ h \in \Oc_{m-1}\} \cup \{ \yv_{r,0}^{m-1} \notin \Uc_{m-1}\} \right \} ,
 \left \{ h \notin \Oc_{m-1} \right \} ,  \left \{ \yv_{r,0}^{m} \notin \Uc_m \right \} \right ) \nonumber \\
& \leq & P \left ( \{ h \in \Oc_{m-1} \}, \{ h \notin \Oc_m \} \right ) +
P \left ( \{ h \notin \Oc_{m-1} \} , \{ \yv_{r,0}^{m-1} \notin \Uc_{m-1} \} \right ),
\end{eqnarray}
where the second inequality follows from the fact that for events
$A,B,C$ and $D$, we have using the distributive law and the union
bound that
\begin{eqnarray*}
P \left ( \{ A \cup B \} \cap \{ C \cap D \} \right ) &=& P \left ( \{ A
\cup (B \cap \overline{A}) \}  \cap \{ C \cap D \} \right )\\
&\leq& P ( A \cap C ) + P ( B \cap \overline{A} ).
\end{eqnarray*}
Finally, for $m = M$, we have
\begin{eqnarray} \label{cavoliM}
P(\Mc = M) & = & P \left ( \overline{\Dc}_1, \ldots, \overline{\Dc}_{M-1} \right ) \nonumber \\
& \leq & P \left ( \overline{\Dc}_{M-1} \right ) \nonumber \\
&=& P \left ( \{ h \notin \Oc_{M-1} \} , \{ \yv_{r,0}^{M-1} \notin \Uc_{M-1} \} \right ) + P \left (h \in \Oc_{M-1} \right ).
\end{eqnarray}
We notice that the event $\{ h
\in \Oc_{m-1} \} \cap \{ h \notin \Oc_m \}$ coincides with
(\ref{m-event}) and therefore the first term in (\ref{cavolim1})
decreases as $\rho^{-\overline{d}_m(r)}$. It is also immediate to
see that $P \left (h \in \Oc_{M-1} \right ) \; \doteq \;
\rho^{-\overline{d}_M(r)}$. Hence, we are left with the analysis of
the probability
\begin{equation} \label{cavolim2}
P \left ( \{ h \notin \Oc_{m} \} , \{ \yv_{r,0}^{m} \notin \Uc_{m}
\} \right )
\end{equation}
for all $ m = 1,\ldots,M-1$.
Averaging with respect to the random coding ensemble, we may choose without loss of generality
$\omega = 1$ as the reference transmitted message.
We have
\[ \overline{\Uc}_m \subseteq \left \{ |\vv_0^{m}|^2  > mT(1 + \delta) \sigma_v^2 \right \} \cup \Rc_m(1), \]
where $\Rc_m(1)$ are the points $\yv_{r,0}^{m}$ such that
$|\yv_{r,0}^{m} - h \xv_{s,0}^{m}(1)|^2  \leq mT(1 + \delta)
\sigma_v^2$, and there exists some $\omega \neq 1$ for which also
$|\yv_{r,0}^{m} - h \xv_{s,0}^{m}(\omega)|^2  \leq mT(1 + \delta)
\sigma_v^2$. Letting for brevity $\Delta\xv(\omega) =
\xv_{s,0}^{m}(\omega) - \xv_{s,0}^{m}(1)$, we can write
\begin{eqnarray*}
\Rc_m(1) & = & \bigcup_{\omega \neq 1} \left \{ \left |\vv_0^{m} -
h \Delta \xv(\omega) \right |^2  \leq mT(1 + \delta) \sigma_v^2, \; \left |\vv_0^{m} \right |^2 \leq mT(1 + \delta)
\sigma_v^2 \right \}.
\end{eqnarray*}
Using the union bound and the Chernoff bound we have
\begin{eqnarray} \label{cavolim3}
P \left ( \{ h \notin \Oc_{m} \} , \{ \yv_{r,0}^{m} \notin \Uc_{m} \} \right ) & \leq &
P \left ( |\vv_0^{m}|^2 \geq mT(1 + \delta)\sigma_v^2 \right )  + P \left ( \{ h \notin \Oc_{m} \}, \Rc_m(1) \right ) \nonumber \\
& \leq &  (1 + \delta)^{mT} e^{-mT\delta}  + \sum_{\omega \neq 1} P
\left ( \left \{h \notin \Oc_m \right \} ,\left \{ \left |\vv_0^{m}
\right |^2 \leq mT(1 + \delta) \sigma_v^2 \right \},
\right . \nonumber \\
& &  \left . \left \{ \left |\vv_0^{m} - h \Delta \xv(\omega) \right
|^2  \leq mT(1 + \delta) \sigma_v^2 \right \} \right )
\end{eqnarray}
Let us consider one term in the sum in the last line of (\ref{cavolim3})
for a given message $\omega$ and given channel $h$, averaged over the random coding
ensemble. Noticing that for vectors $\av$ and $\bv$ and $\Gamma > 0$
we have
\[ \{|\av + \bv|^2 \leq \Gamma, |\bv|^2 \leq \Gamma \} \subseteq \{|\av|^2 \leq 4\Gamma \}, \]
we can bound this probability as
\begin{eqnarray} \label{mimoarq}
& P \left ( \left . \left \{ \left |\vv_0^{m} - h \Delta\xv(\omega)
\right |^2  \leq mT(1 + \delta) \sigma_v^2 \right \}, \left\{ \left
|\vv_0^{m} \right |^2 \leq mT(1 + \delta) \sigma_v^2 \right \}
\right | h \right )& \\
& \leq P \left ( \left . \left |h \Delta \xv(\omega) \right |^2 \leq 4 mT(1 + \delta) \sigma_v^2 \right | h \right )& \nonumber \\
& \stackrel{(a)}{=} P \left ( \left . \rho' |h|^2 \chi \leq 2 mT(1 + \delta) \right | h \right )& \nonumber \\
& \dot\leq \rho^{-mT [1 - \nu]_+},&
\end{eqnarray}
where (a) follows from the fact that for the randomly generated
codewords, $\chi = |\Delta\xv(\omega)|^2/E$ is a central chi-squared
random variable with mean $2mT$ and $2mT$ degrees of freedom, and
the last line follows by letting $|h|^2 = \rho^{-\nu}$ and from the
fact that the chi-squared cdf satisfies $P(\chi \leq u) =
O(u^{mT})$ for small $u$ and $P(\chi \leq u) = O(1)$ for large
$u$. Summing over the $\rho^{rMT} - 1$ messages $\omega \neq 1$ and
integrating with respect to the pdf of $|h|^2$ over the set
$\overline{\Oc}_m$, we obtain
\begin{eqnarray} \label{cavolim4}
P \left ( \{ h \notin \Oc_{m} \} , \{ \yv_{r,0}^{m} \notin \Uc_{m} \} \right ) & \dot\leq &
\int_{\{\nu \geq 0, [1 - \nu]_+ \geq \frac{Mr}{m}\}} \rho^{-\nu} \;\;   \rho^{-mT[1 - \nu]_+ + rMT} \; d\nu  \nonumber \\
& \doteq  & \rho^{-\widetilde{d}_m(r)},
\end{eqnarray}
where, from a standard application of Varadhan's lemma, we have
\begin{equation} \label{cavoli-exponent}
\widetilde{d}_m(r) = \inf_{\nu \geq 0, [1 - \nu]_+ \geq
\frac{Mr}{m}} \; \left \{ \nu + mT [ 1 - \nu]_+ - rMT \right \}.
\end{equation}
The domain of $\nu$ over which the infimum is calculated is
non-empty only for $r \leq \frac{m}{M}$. This means that the set of
channels for which the probability in (\ref{cavolim2}) has a
polynomial decrease is empty for $r > \frac{m}{M}$ and therefore
$\widetilde{d}_m(r) = \infty$ for $r > \frac{m}{M}$. For $r \leq
\frac{m}{M}$ it is not hard to see that for all $T \geq 1$ we have
$\widetilde{d}_m(r) = 1 - \frac{Mr}{m}$. Comparing
$\widetilde{d}_m(r)$ with $\overline{d}_m(r)$ we see that the former
dominates the latter for all $r \in [0,1]$. It follows that for our
relay bounded distance decoder and the Gaussian random coding
ensemble $P(\Mc = m) \; \dot\leq \; \rho^{-\overline{d}_m(r)}$.

So far we have shown that in the upper bound (\ref{cond-error}) the
terms $P (\Ec_r |\Mc = m)$ are asymptotically negligible and the
terms $P (\Mc = m )$ are upper bounded by the same exponent of the
outage probability based, infinite $T$, case. It remains to show
that the terms $P (\Ec |\overline{\Ec}_r,\Mc = m)$ have exponent
$d_m(r)$ given in \eqref{dm1}, \eqref{dm2}, and the proof will be
complete.

We consider the GLRT decoder at the destination. This decoder
ignores the knowledge of the a priori distribution of $\Mc$ and
treats it as a deterministic unknown parameter. Hence, we are in the
presence of a compound channel formed by the family of $m$-SC
component channels, without any a priori knowledge of $\Mc$.

Again, without loss of generality we assume message $1$ is
transmitted. While for the sake of notational simplicity, we omit
the explicit conditioning with respect to $\overline{\Ec}_r$, it is
understood that the relay has perfect knowledge of the transmitted
information message. We omit also the explicit conditioning with
respect to CSIR and denote $\yv_{s,0}^{M}$ simply by $\yv$ since no
ambiguity is possible at this point. Hence, the likelihood function
$p\left (\yv_0^{M} | \omega ,m,g_1,g_2 \right )$ shall be denoted
simply by $p(\yv|\omega,m)$. The pairwise error probability for some
$\omega \neq 1$ can be upper bounded as follows:
\begin{eqnarray}
P (1 \rightarrow \omega | \Mc = m)
& = & P \left ( \left . \max_{m'} p(\yv|1,m') \leq  \max_{m'} p(\yv|\omega,m') \right | \Mc = m \right ) \nonumber \\
& \leq & P \left ( \left .  p(\yv|1,m) \leq  \max_{m'} p(\yv|\omega,m') \right  | \Mc = m \right ) \nonumber \\
& = & P \left ( \left .  \bigcup_{m'=1}^M \left \{ p(\yv|1,m) \leq  p(\yv|\omega,m') \right \} \right  | \Mc = m \right ) \nonumber \\
& \leq & \sum_{m'=1}^M P \left ( \left .  p(\yv|1,m) \leq   p(\yv|\omega,m') \right  | \Mc = m \right ).
\end{eqnarray}
We shall analyze separately the terms inside the above sum, averaged over the random coding ensemble.
Define the event
\[ \Ec_1 = \left \{  \frac{p(\yv|\omega,m')}{p(\yv|1,m)} \geq 1 \right \}.  \]
We first analyze the probability of the event $\Ec_1$, which we then
use to compute $P(\Ec)$. Assuming $\Mc = m$, the actual received
signal is
\begin{eqnarray}
\yv_0^{m} & = & g_1 \xv_{s,0}^{m}(1) + \wv_0^{m} \nonumber \\
\yv_{m}^{M} &=& g_1 \xv_{s,m}^{M}(1)  + g_2 \xv_{r,m}^{M}(1) +
\wv_{m}^{M}.
\end{eqnarray}
We consider the case $m' \geq m$
and leave the case $m' \leq m$ to the reader, since it follows in an almost identical manner.
Define the partial codeword differences
$\Delta \xv_{s,0}^m = \xv_{s,0}^{m}(1) - \xv_{s,0}^{m}(\omega)$,
$\Delta \xv_{s,m}^{m'} = \xv_{s,m}^{m'}(1) - \xv_{s,m}^{m'}(\omega)$,
$\Delta \xv_{s,m'}^{M} = \xv_{s,m'}^{M}(1) - \xv_{s,m'}^{M}(\omega)$, and
$\Delta \xv_{r,m'}^{M} = \xv_{r,m'}^{M}(1) - \xv_{r,m'}^{M}(\omega)$.
The error event $\Ec_1$ can be written as
\begin{eqnarray}
\Ec_1 & = & \left \{ \phantom{\sum}  |g_1|^2 \left |\Delta \xv_{s,0}^m \right |^2 +
2 \Re \left \{ g_1 (\wv_0^{m})^\herm  \Delta \xv_{s,0}^m \right \} +  \left | g_1 \Delta\xv_{s,m}^{m'} + g_2 \xv_{r,m}^{m'}(1) \right |^2 + \right . \nonumber \\
& & + 2 \Re \left \{ ( \wv_{m}^{m'} )^\herm \left [ g_1 \Delta\xv_{s,m}^{m'} +  g_2 \xv_{r,m}^{m'}(1) \right ] \right \} +
\left |g_1 \Delta \xv_{s,m'}^M  + g_2 \Delta \xv_{r,m'}^M  \right |^2 + \nonumber \\
& & \left . + 2 \Re\left \{ (\wv_{m'}^{M})^\herm \left [ g_1 \Delta \xv_{s,m'}^M +
 g_2 \Delta \xv_{r,m'}^{M} \right ] \right \}  \leq 0 \phantom{\sum} \right
 \}.
\end{eqnarray}
After a little algebra, we obtain the compact expression
\[ \Ec_1 =  \left \{ 2 \Re \{ \zv^\herm \wv \} \leq -|\zv|^2 \right \}, \]
where $\zv$ is defined as
\[ \zv \triangleq \left [
\begin{array}{c}
g_1 \Delta\xv_{s,0}^m \\
g_1 \Delta\xv_{s,m}^{m'} + g_2 \xv_{r,m}^{m'} \\
g_1 \Delta_{s,m'}^M + g_2 \Delta \xv_{r,m'}^M \end{array} \right ] \]
For given codebooks $\Xc_s,\Xc_r$, the variance of $2 \Re \{ \zv^\herm \wv \}$ is equal to
$2|\zv|^2\sigma_v^2$, which leads to
\[ P(\Ec_1| \Xc_s,\Xc_r, \Mc=m,g_1,g_2) \leq Q \left( \frac{|\zv|}{\sqrt{2 \sigma_v^2}} \right) \leq e^{-|\zv|^2/(4\sigma_v^2)}. \]
Define the following notation,
\[ \xiv_i = [x_{s,i}(1) \ x_{s,i}(\omega) \ x_{r,i}(1) \ x_{r,i}(\omega)]^\transp, \ 1 \leq i \leq MT, \]
and
\[ \xiv \triangleq [\xiv_1^\transp \xiv_2^\transp \cdots \xiv_{MT}^\transp]^\transp \in
\mathcal{C}^{4MT \times 1}. \]
It can be verified that $|\zv|^2 = \xiv^\herm \Mm \xiv$, for a block diagonal $\Mm$ of the form
\[
\Mm = \left[
\begin{array}{ccc}
\Mm_1 & & \\
 & \ddots & \\
 & & \Mm_{MT}
\end{array}
\right],
\]
where for $\ 1 \leq k \leq mT$,
\[ \Mm_k = |g_1|^2 \left[
     \begin{array}{cccc}
       1 & -1 & 0 & 0 \\
       -1 & 1 & 0 & 0 \\
       0 & 0 & 0 & 0 \\
       0 & 0 & 0 & 0
     \end{array}
   \right],
\]
for $mT+1 \leq k \leq m'T$,
\[
   \Mm_k = \left[
     \begin{array}{cccc}
       |g_1|^2 & -|g_1|^2 & g_2 g_1^* & 0 \\
       -|g_1|^2 & |g_1|^2 & -g_2 g_1^* & 0 \\
       g_1 g_2^* & -g_1 g_2^* & |g_2|^2 & 0 \\
       0 & 0 & 0 & 0
     \end{array}
   \right],
\]
and for $m'T+1 \leq k \leq MT$,
\[
   \Mm_k = \left[
     \begin{array}{cccc}
       |g_1|^2 & -|g_1|^2 & g_2 g_1^* & -g_2 g_1^* \\
       -|g_1|^2 & |g_1|^2 & -g_2 g_1^* & g_2 g_1^* \\
       g_1 g_2^* & -g_1 g_2^* & |g_2|^2 & -|g_2|^2 \\
       -g_1 g_2^* & g_1 g_2^* & -|g_2|^2 & |g_2|^2
     \end{array}
   \right].
\]
It turns out that the matrices $\Mm_k$ have rank $1$, for all $1
\leq k \leq MT$. It follows that the eigenvalues of each $\Mm_k$ are
$\trace (\Mm_k), 0,0,0$. We now average $P(\Ec_1|\Xc_s,\Xc_r,
m,g_1,g_2)$ over the ensemble of random Gaussian codebooks. In order
to do so, we use the following well-known result on the
characteristic function of Hermitian quadratic form of complex
Gaussian random variables (briefly, HQF-GRV).
\begin{lem} \cite[Appendix 4]{SchBenSte}
The characteristic function of the HQF-GRV $\Delta = \zv^\herm \Fm
\zv$, where $\zv \sim \Cc\Nc(\overline{\zv},\Rm)$ is given by
\[ \Phi_\Delta (s) = \mathbb{E}\left[ \exp (-s \Delta) \right] =
\frac{\exp (-s \overline{\zv}^\herm \Fm (\Id + s \Rm \Fm)^{-1}
\overline{\zv})}{\det (\Id + s \Rm \Fm)}. \]
\end{lem}
Therefore,
\begin{eqnarray*}
P(\Ec_1|m,g_1,g_2) & \leq & \EE_{\Xc_s,\Xc_r} \left [ e^{-|\zv|^2/(4\sigma_v^2)} \right ] \nonumber \\
& = & \Phi_{|\zv|^2} \left ( \frac{1}{4\sigma_v^2} \right ) \nonumber \\
& = & \frac{1}{\det (\Id + \frac{\rho}{4} \Mm)}.
\end{eqnarray*}
Explicitly, we have
\begin{eqnarray} \label{eq:pE1}
\frac{1}{\det (\Id + \frac{\rho}{4} \Mm)} &=& \frac{1}{ \left[ 1 + \frac{\rho}{2} |g_1|^2 \right]^{mT}}
 \cdot \frac{1}{\left[ 1 + \frac{\rho}{4} (2|g_1|^2 + |g_2|^2)\right]^{(m'-m)T}}
 \cdot \frac{1}{\left[ 1 + \frac{\rho}{2} ( |g_1|^2 + |g_2|^2) \right]^{(M-m')T}}\nonumber\\
&\doteq& \frac{1}{ \left[ 1 + \rho |g_1|^2 \right]^{mT}} \cdot
\frac{1}{\left[ 1 + \rho (|g_1|^2 + |g_2|^2)\right]^{(M-m)T}}.
\end{eqnarray}
We notice that (\ref{eq:pE1}) does not depend on $m'$, at least in
the exponential equality sense. Summing over all $m' = 1,\ldots,M$
and over all messages $\omega \neq 1$, we eventually can bound the
average probability of error of the GLRT decoder conditioned on $\Mc
= m$ and on $g_1,g_2$ as
\begin{eqnarray} \label{wow}
P (\Ec |\overline{\Ec}_r,\Mc = m,g_1,g_2) & \leq & \sum_{\omega\neq 1} P (1 \rightarrow \omega | \Mc = m, g_1,g_2) \nonumber \\
& \leq & \sum_{\omega \neq 1} \sum_{m'=1}^M P(\Ec_1|m,g_1,g_2) \nonumber \\
& \dot\leq & \frac{M \rho^{rMT}}{ \left[ 1 + \rho |g_1|^2 \right]^{mT}} \cdot \frac{1}{\left[ 1 + \rho (|g_1|^2 +
|g_2|^2)\right]^{(M-m)T}}.
\end{eqnarray}
Next, we shall evaluate the diversity exponent of $P (\Ec
|\overline{\Ec}_r,\Mc = m) = \EE_{g_1,g_2}[ P (\Ec
|\overline{\Ec}_r,\Mc = m,g_1,g_2) ]$. In order to do so, we
separate the outage event from the no-outage event. Define the
outage event of the $m$-SC as
\begin{eqnarray}
\Ac_{m} & = & \big\{ (M-m) \max \{ [1-\alpha_1]_+, [1-\alpha_2]_+ \} + m [1-\alpha_1]_+ - rM \leq 0 \big\}.
\end{eqnarray}
Then,
\begin{eqnarray}
P(\Ec|\overline{\Ec}_r,\Mc = m) & = & P ( \Ec, \overline{\Ac}_{m} |\overline{\Ec}_r,\Mc = m)  + P ( \Ec, \Ac_{m} |\overline{\Ec}_r,\Mc = m) \nonumber \\
&\leq& P (\Ac_m) + P (\Ec, \overline{\Ac}_{m}|\overline{\Ec}_r,\Mc = m).  \nonumber \\
& &
\end{eqnarray}
Recall that
\[ P(\Ac_m) = P_{out}^{m-\text{SC}}(r) \doteq \rho^{-d_{m}(r)}, \]
where $d_{m}(r)$ is evaluated in \eqref{dm1}, \eqref{dm2}. In order
to evaluate $P(\Ec , \overline{\Ac}_m| \overline{\Ec}_r, \Mc = m)$, we use
(\ref{wow}) and write the exponential inequality
\[ P (\Ec |\overline{\Ec}_r,\Mc = m,g_1,g_2) \; \dot\leq \; \rho^{- T g_m(\alpha_1,\alpha_2,r)}, \]
where
\begin{eqnarray} \label{gg}
g_{m}(\alpha_1,\alpha_2,r) & = & (M-m) \max \{ [1-\alpha_1]_+, [1-\alpha_2]_+ \} + m[1-\alpha_1]_+ - rM.
\end{eqnarray}
Therefore, using again Varadhan's lemma, we obtain
\[ P(\Ec, \overline{\Ac}_m|\overline{\Ec}_r,\Mc = m) \; \doteq \; \rho^{-d_{G,m}(r)},\]
where
\begin{equation} \label{wow1}
d_{G,m}(r) = \inf \limits_{
\begin{array}{c}
g_{m}(\alpha_1,\alpha_2,r) > 0 \\
\alpha_1,\alpha_2 \geq 0 \end{array}
} \left \{ \alpha_1 + \alpha_2 + T g_{m}(\alpha_1,\alpha_2,r) \right \}.
\end{equation}
The above infimum is achieved when $g_{m}(\alpha_1,\alpha_2,r) \downarrow 0$, yielding
\[ d_{G,m}(r) = d_{m}(r).\]
This concludes the proof of Theorem \ref{achievability-thm}.
\end{proof}

{\bf Remark.} The proof of Theorem \ref{achievability-thm} is not
only conceptually appealing, but also reveals a few very important
and often neglected features that  should be taken into account in
the design of a DDF scheme. First, the proof sheds light on the fact
that the relay must make its decision based not only on the outage
condition, but also on the reliability of the decoding decision.
Then, it shows also that despite the fact that the destination does
not know the relay decision time, there is no need for an explicit
protocol that provides this side information. In
Appendix~\ref{app:RAD}, we analyze a simpler decoder, nicknamed {\em
relay activity detector}, based on separated detection of the  relay
decision time by treating the codewords as random Gaussian signals
(i.e., ignoring the structure of the code). We show that such a
simple ``energy detector'' is optimal if we let $T \rightarrow
\infty$ first, and then consider the high SNR performance, but it is
dramatically suboptimal if we do the limits in the reverse order. In
fact, for any finite $T$, the relay activity detector yields a
constant error probability, that does not vanish with SNR.

\subsection{Computing the DMT and comparisons \label{sec:pareto}}

\begin{figure}[h]
\begin{center}
\includegraphics[width=15cm]{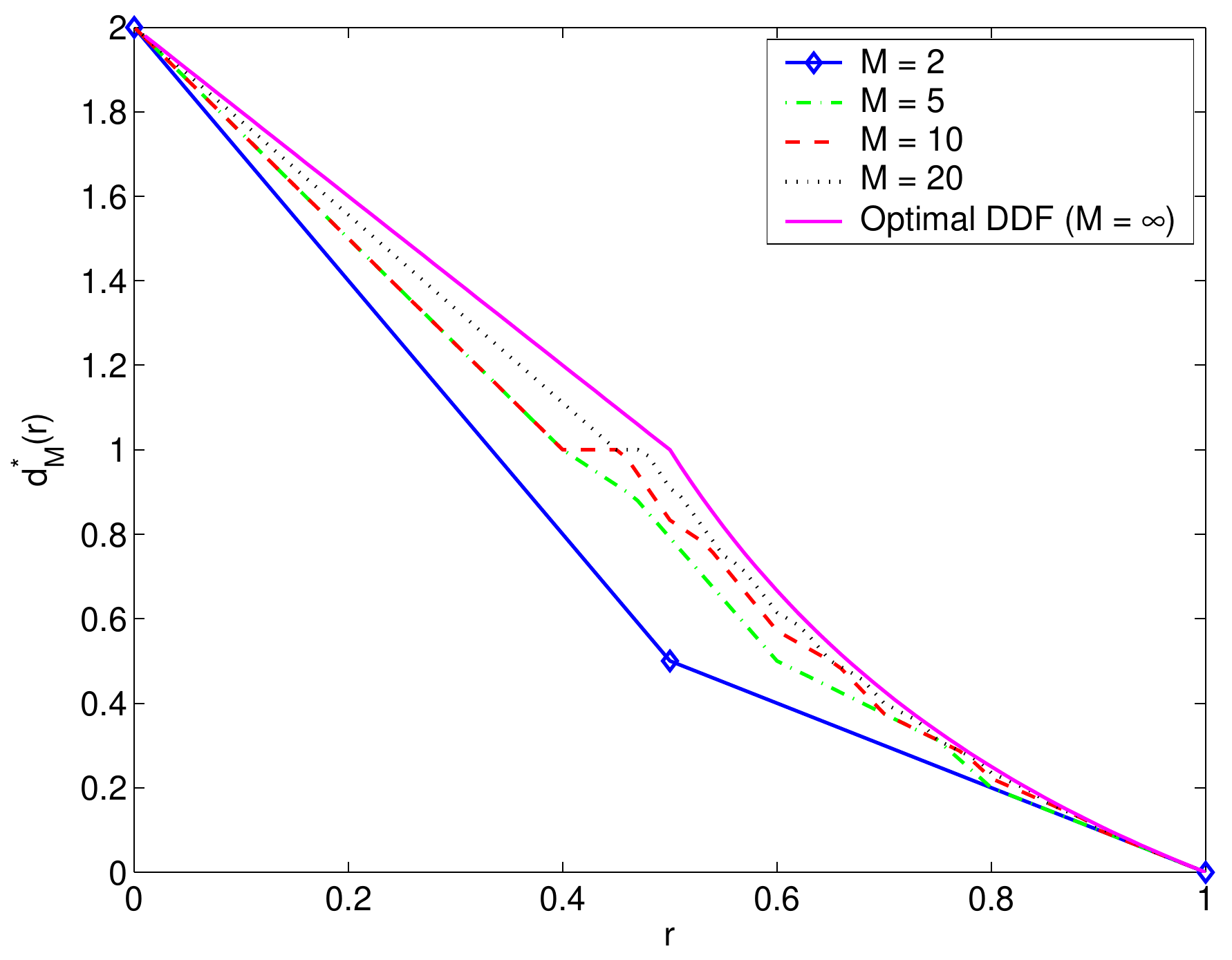}
\caption{The DMT of the DDF channel with finitely many decoding decision times.
\label{fig:Ala_DDF}}
\end{center}
\end{figure}

Obtaining a closed-form solution to the DMT expression in
Theorem~\ref{th:DMT_DDF} appears to be intractable. We plot in
Fig.~\ref{fig:Ala_DDF} values of $d^*_M(r)$ for $M = 2,5,10$ and
$20$ in comparison with the optimal DMT of the DDF protocol
(corresponding to $M = \infty$).

With increasing $M$, $d^*_M(r)$ is seen to approach the optimal
tradeoff very rapidly. For practical code design, even a relatively small
value of $M$ is therefore expected to have close to optimal performance in
terms of diversity.

{\em Remark.} The authors in \cite{MurAzaGam} consider a related
problem, where $T \rightarrow \infty$ and the relay is restricted to
a finite number of decision times (say $N$). These time instants
coincide with the end of blocks $\{M_j\}_{j=1}^{N}$, with $1 \leq
M_1 < \cdots < M_N < M \ \forall \ j$ (notice: with this notation,
in our case we would have $N = M$ and $M_j = j$). Further, define
$M_0 \triangleq 0, \ M_{N+1} \triangleq M$, and a set of ``waiting
fractions'' $\{ f_j \}_{j=0}^{N+1}$ by $f_j \triangleq
\frac{M_j}{M}$. Thus
\[ f_0 = 0 < f_1 < \cdots < f_N < f_{N+1} = 1. \]
In \cite{MurAzaGam}, it is proved that for any fixed $N$ no set of waiting fractions yields a DMT curve that dominates all others.
Then, a particular set of waiting fractions are chosen that yield for any fixed $N$ a DMT curve
that is not uniformly dominated~\footnote{According to the definition in \cite{MurAzaGam},
protocol A {\em uniformly dominates} protocol B if, for any
multiplexing gain $r$, $d_A(r) \geq d_B(r)$. A protocol that is not uniformly dominated by any other protocol is said to be Pareto-optimal.}
by any other protocol with the same number of decision times $N$.
The resulting DMT is derived and it is summarized by the following lemma from \cite{MurAzaGam}.
\begin{lem} \cite{MurAzaGam}
For the DDF protocol with a given number $N$ of decision times, let $f_1^p = \frac{1}{2}$ and
\[ f_j^p = \frac{1 - f^p_{j-1}}{2 - \left(1 + \frac{1}{f^p_N} \right)f^p_{j-1}} , \text{ for } 1< j \leq N, \]  then
no set of fractions uniformly dominates $\{f^p_j\}_{j =1}^N$.
Further, the DMT corresponding to the set of fractions $\{f^p_j\}_{j
=1}^N$ is given by
\begin{equation} \label{eq:DMT_pareto}
d^p(r) = 1 - r + \left [ 1 - \frac{r}{f^p_N} \right ]_+.
\end{equation}
\end{lem}
A few interesting observations can be made about this result. As it is
remarked in \cite{MurAzaGam}, the DMT obtained through $\{f_j^p\}$
is not asymptotically optimal, i.e., it does not converge to the
optimal DMT of the DDF protocol as $N \rightarrow \infty$. Indeed,
it is evident from \eqref{eq:DMT_pareto} that $d_p(r)$ consists of
two straight line segments, say $\mathcal{L}_1$ for $0 \leq r \leq
f_N^p$ and $\mathcal{L}_2$ for $f_N^p \leq r \leq 1$. As $N
\rightarrow \infty$, $\mathcal{L}_1$ and $\mathcal{L}_2$ can at best
be tangential to the curved part of the DMT of the DDF protocol
(i.e., the $0.5 \leq r \leq 1$ region) in \eqref{eq:DMT_DDF}. In
particular, the optimal value of the DMT of the DDF protocol
$d^*(0.5) = 1$ is never approached even in the limit by $d^p(r)$. In
contrast, the DMT $d^*_M(r)$ derived in this paper is asymptotically
optimal. As the number of decoding points increases, $d^*_M(r)$
dominates over $d^p(r)$ for almost all values of $r$, and is
strictly less for only an exceedingly small range of values of $r$.
Asymptotically, it is clear that the only set of points where
$d^p(r)$ dominates $d^*_M(r)$ is a very small set of points around
the point where $d^p(r)$ is tangent to the curved part of the DMT of
the DDF protocol.

\section{DMT optimal codes for the single relay DDF channel} \label{code-construction.sect}

The authors in \cite{AzaGamSch} used the ensemble of random Gaussian
codes of asymptotically large block-lengths to show the
achievability of the DMT of the DDF protocol. Subsequently, a
construction of codes derived from cyclic division algebras (CDA)
was shown to achieve the DMT of the DDF channel for arbitrary number
of relays \cite{EliKum}; i.e., they achieve the corresponding
tradeoff for a particular number of decoding instants. As we
increase the block-length and the number of decoding instants, the
DMT of these codes tends towards the optimal DMT of the DDF protocol
given in \eqref{eq:DMT_DDF}.
In a recent submission, \cite{EliKum_arxiv}, the authors present a division-algebraic
construction based on the Alamouti code that is similar in flavor to
the construction to be presented in this paper. However, for the
codes in \cite{EliKum_arxiv}, the parameter $T$ is fixed to $2$; on
the contrary, we will see that our code construction is valid for
arbitrary values of $T$ including the special case of $T = 1$, and
is hence a minimum delay construction. Decoding these codes involves
sphere or sequential decoding \cite{DamGamCai,MurGamDamCai} over a
large dimensional lattice. It is hence of interest to construct
codes that achieve the DMT of the DDF protocol and permit low
complexity decoding. Since our construction is of minimum delay, the
dimensionality of the lattice to be sphere decoded is half that of
the corresponding case in \cite{EliKum_arxiv}.

In order to completely specify a signalling scheme $(\mathcal{X}_s,
\phi, \mathcal{X}_r)$ for the DDF channel, we need to define the
following:
\begin{enumerate}
\item A code $\mathcal{X}_s$ that is used by the source.

\item A causal
{\em decoding decision function} $\phi(\cdot,\cdot):
(\mathbb{C},\mathbb{C}^{MT}) \rightarrow \{ 1,2,\hdots,M \}$, that
dictates when the relay attempts to decode the source's transmission
based on the S-R channel gain $h$ and the signal $\yv_r$ received at
the relay. In particular, if $\phi(h,\yv_r) = M$, the relay will not
attempt to decode the transmission of the source. If $\phi(h,\yv_r)
= m, \ 1 \leq m < M$, then the relay attempts to decode the
transmission of the source upon completion of the $m^\text{th}$
block. Because of the causality constraint, we assume that the
output of $\phi$ at time $m$ depends only on $h$ (CSIR of the relay)
and on $\yv_{r,0}^{m}$.

\item A code $\mathcal{X}_r$ used by the relay. In
the following, we will only consider the case that the relay
implements the Alamouti-DDF scheme \cite{MurAzaGam} given in
\eqref{eq:Ala_scheme}; hence $\mathcal{X}_r$ is the same as
$\mathcal{X}_s$ upto coordinate permutations, sign change and
conjugation.
\end{enumerate}

\subsection{Design tradeoffs}

Despite the importance of the decoder at the destination, as
evidenced in the proof of Theorem 2, in this section we take a
shortcut and we do not treat the the GLRT decoder explicitly. For
the sake of simplicity, our simulations assume a genie-aided
destination, ideally informed of the relay decision time. The main
focus of this section is on the design of the codebook $\Xc_s$ and
on the efficient implementation of the relay decoding decision
function, in order to trigger the relay transmission only when
decisions are reliable.

Choosing a good relay decoding decision decision function $\phi$ is
critical to ensure good performance: a conservative $\phi$ that
makes the relay wait for too long before decoding results in low
relay error probability $P(\mathcal{E}_r)$, but increases the
destination error probability
$P(\mathcal{E},\overline{\mathcal{E}}_r)$  since the relay has less
time to help the destination. Vice-versa, a $\phi$ that is too
aggressive and makes the relay decode too early yields low
$P(\mathcal{E},\overline{\mathcal{E}}_r)$ but results in a large
$P(\mathcal{E}_r)$, since the blocklength of the signal observed at
the relay is too short to cope with atypical noise. We shall also
see through simulations that undetected decoding errors at the relay
have a huge impact on performance, since the relay ends up jamming
the destination with high probability. We will present our choices
of $\mathcal{X}_s$ and $\phi$ in the following two subsections.

\subsection{Approximately universal $\mathcal{X}_s$}

The equivalent channel resulting from the use of the Alamouti
scheme for the relay code is a parallel channel
\eqref{eq:AlaDDF_channel} with statistically dependent fading coefficients.
We will choose $\mathcal{X}_s$ to be a code of length $MT$ that is
{\em approximately universal} over the parallel fading channel. A
code that is approximately universal over the parallel channel (a notion
introduced in \cite{TavVis_IT}) meets the DMT
over {\em any} parallel channel. Such a code has an error
probability that decays exponentially with $\rho$ for all parallel channel gains such that
the corresponding mutual information is larger than the coding rate, i.e.,
for all channel gains in the no-outage region. Therefore, such an
approximately universal code $\mathcal{X}_s$ meets the DMT of the
relay DDF channel for any $M$. This means that, for any fixed rate $R$ and sufficiently
large SNR, the decay of error probability with SNR of our code
(with finite $T$) exhibits the same slope of outage probability. However, the ``gap from outage'' (i.e., the horizontal distance in dB between the outage probability and the actual probability of error) is not captured by the DMT optimality and in practice
it may be very large, thus making a DMT-optimal scheme totally useless for practical purposes.
We shall discuss ways to close this gap in the next section, by an appropriate choice of the relay decision function $\phi$.

We may obtain approximately universal $\mathcal{X}_s$ from either
suitable algebraic lattices \cite{BayOggVit,VitOgg} or from
permutation codes through UDMs \cite{TavVis_IT,GanVon}.
In the following we briefly review these constructions.

\subsubsection{Rotated QAM codes from algebraic lattices}
Let $\mathbb{L}$ be an $MT$-dimensional extension of
$\mathbb{Q}(\imath)$ and let the Galois group
$Gal(\mathbb{L}|\mathbb{Q}(\imath)) = \{ \sigma_1, \hdots,
\sigma_{MT} \}$. Denote the ring of integers of $\mathbb{L}$ as
$\mathcal{O}_{\mathbb{L}}$ and let $\mathcal{I}$ be an ideal of
$\mathcal{O}_{\mathbb{L}}$. Let $N_{\mathbb{L}|\mathbb{Q}(\imath)}
(\cdot)$ denote the algebraic norm from $\mathbb{L}$ to
$\mathbb{Q}(\imath)$. We define the code $\mathcal{X}_s$ as follows:
\[ \mathcal{X}_s = \left\{ \left. \left[
\begin{array}{c}
\sigma_1(\ell)\\ \sigma_2(\ell)\\ \vdots \\ \sigma_{MT}(\ell)
\end{array}
\right] \right| \ell \in \mathcal{S} \right\}, \] where
$\mathcal{S}$ is some finite subset of $\mathcal{I}$.
$\mathcal{X}_s$ has the desirable property of a ``non-vanishing''
product distance, since we have for each $\xv \in \mathcal{X}_s$
that
\begin{eqnarray*}
\prod_{j=1}^{MT} |x_j| = \left | \prod_{j=1}^{MT} \sigma_j(\ell) \right |
= \left | N_{\mathbb{L}|\mathbb{Q}(\imath)}(\ell) \right |  \geq 1,
\end{eqnarray*}
since the norm $N_{\mathbb{L}|\mathbb{Q}(\imath)}(\cdot)$ of an
algebraic integer in $\mathbb{L}$ is an element of
$\mathbb{Z}[\imath]$. This non-vanishing product distance property
ensures that $\mathcal{X}_s$ is approximately universal over the
parallel channel \cite{TavVis_IT,EliKum_arxiv}.

It can be verified that $\mathcal{X}_s$ can equivalently be
rewritten as a lattice code, i.e.,
\begin{equation} \label{eq:code_lattice}
\mathcal{X}_s = \left\{ \left. \Gm \bv \right| \bv \in \mathcal{B}
\right\},
\end{equation} for suitable $\Gm \in \mathbb{C}^{MT \times
MT}$ and $\mathcal{B} \subset \mathbb{Z}[\imath]^{MT}$. A particular
choice of $\Gm$ and $\mathcal{B}$ that is good in terms of shaping
consists of constructing $\Gm$ to be unitary and $\mathcal{B}$ to be
a set of points in $\mathbb{Z}[\imath]^{MT}$ contained in a
hypercube that is centered around the origin~\footnote{Notice
however that choosing $\Gm$ unitary is optimal only when we are
constrained to use a {\em linear} map to encode the information
vector onto the code symbols. An alternate approach is to use a
constellation carved out of a dense lattice in $\mathbb{R}^n$ and
employ a {\em non-linear} sphere encoder and a mod-$\Lambda$
MMSE-GDFE lattice decoder; this has been shown to yield significant
performance improvements over unitary shaping
\cite{KumCai_ISIT06,KumCai_ISIT07}. For simplicity of exposition, we
will restrict our attention to the case of linear encoding in this
paper.}. For the algebraic details regarding the construction of
such unitary $\Gm$, see \cite{BayOggVit,VitOgg}. Notice also that
choosing the information set $\Bc$ to be a bounded subset of
$\ZZ[\imath]^{MT}$ corresponds, in practice, to choosing information
symbols from a QAM alphabet, which is appealing for practical
implementation. The rate of $\mathcal{X}_s$ in this case is
\[ R = \frac{\log |\mathcal{B}|}{MT} \text{ bpcu}.\]

\emph{Parameters for simulations:} In the simulations to follow in
Sec.~\ref{sec:simul}, we construct the matrix $\Gm$ using the
cyclotomic construction given in \cite{BayOggVit}. For $M = 4$ and
$T = 1$, $\Gm$ is a complex $4 \times 4$ matrix, or equivalently a
real $8 \times 8$ matrix. We choose $\Bc$ to be a cartesian product
of $Q^2$-QAM alphabets,
\[ \Bc = \left\{ a + \imath b | -Q+1 \leq a,b \leq Q-1, \ a,b \text{ odd} \right\}^{MT},\]
for some even integer $Q$. Thus $|\Bc| = Q^{2MT}$.
For example, by choosing $Q = 4$ with $M = 4$ and $T = 1$ we obtain a rate of
$R = 4$ bpcu.

\subsubsection{Permutation codes from UDM}
Approximately universal code construction from UDM were introduced in \cite{TavVis_IT}
and a general algebraic construction valid for any number of sub-channels was
provided in \cite{GanVon}. \\[12pt]
\begin{defn} \cite{GanVon}
Let $n$ and $L$ be some positive integers and let $q$ be a prime
power. The $L$ matrices $\Am_0,\hdots,\Am_{L-1}$ over $\mathbb{F}_q$
of size $n \times n$ are $(L,n,q)$-UDMs if for every
$(k_0,\hdots,k_{L-1})$ such that $0 \leq k_\ell \leq n \ \forall \
\ell$, $\sum_{\ell = 0}^{L-1} k_\ell \geq n$,
the $(\sum_{\ell = 0}^{L-1} k_\ell) \times n$ matrix composed of the first $k_0$ rows
of $\Am_0$, the first
$k_1$ rows of $\Am_1$, $\hdots$, the first $k_{L-1}$ rows of
$\Am_{L-1}$ has full rank. \hfill $\lozenge$ \\[12pt]
\end{defn}
The authors in \cite{GanVon} provide an algebraic construction of
such $(L,n,q)$-UDMs for any $L \leq q+1$. It is shown in
\cite{TavVis_IT} that an approximately universal permutation code
for the parallel channel with $L$-branches can be obtained from
$(L,n,q)$-UDMs, in the following manner. Assume that we have to
transmit $2n$ information symbols from $\mathbb{F}_q$. We encode
independently $n$-symbols each onto the I and Q sub-channels. Let
$\uv \in \mathbb{F}_q^n$ denote the first $n$ input information
symbols. Map the sequence of $\mathbb{F}_q^n$ symbols $\{ \Am_1 \uv,
\Am_2 \uv, \hdots, \Am_{L} \uv\}$ componentwise onto a $L$-length
vector of $q^n$-PAM symbols, and transmit the components on the I
sub-channel. The next $n$ information symbols are similarly encoded
and transmitted on the Q sub-channel. The rate of such a permutation
code is
\[ R = \frac{2n \log q}{L} \text{ bpcu}.\]
In our case, we set $L = MT$ to obtain codes for the DDF channel.

\emph{Parameters for simulations:} The simulations involving
permutation codes in Sec.~\ref{sec:simul} for $M=4$, $T=1$ are
derived from $(4,4,4)$-UDMs, leading to $R = 4$ bpcu. In order to
completely specify the code, we need to provide the mapping to PAM
symbols that was used. We construct the Galois field $\mathbb{F}_4$
using the primitive polynomial $X^2+X+1$. Thus any element in
$\mathbb{F}_4$ may be associated with a polynomial $b_1 X + b_0$,
where the $b_i$ are either $0$ or $1$, and $X$ is a primitive
element. Hence we may also associate each element in $\FF_4$ with
the binary string $b_1 b_0$. In order to map an $\FF_4^4$ vector
$\vv$ (which is one of the $\Am_j \uv$ considered previously) to the
PAM alphabet, first concatenate the binary strings corresponding to
$v_{i} \in \FF_4$, $i = 1,2,3,4$ to obtain an $8$-length binary
vector $\bv$. This binary vector is mapped to the centered 256-PAM
alphabet by computing $2 \sum_{i=0}^7 b_i 2^i - 255$.

\subsection{Decoding decision function $\phi$ and Forney's decision rule \label{sec:simul}}

\begin{figure}[h]
\begin{center}
\includegraphics[width=15cm]{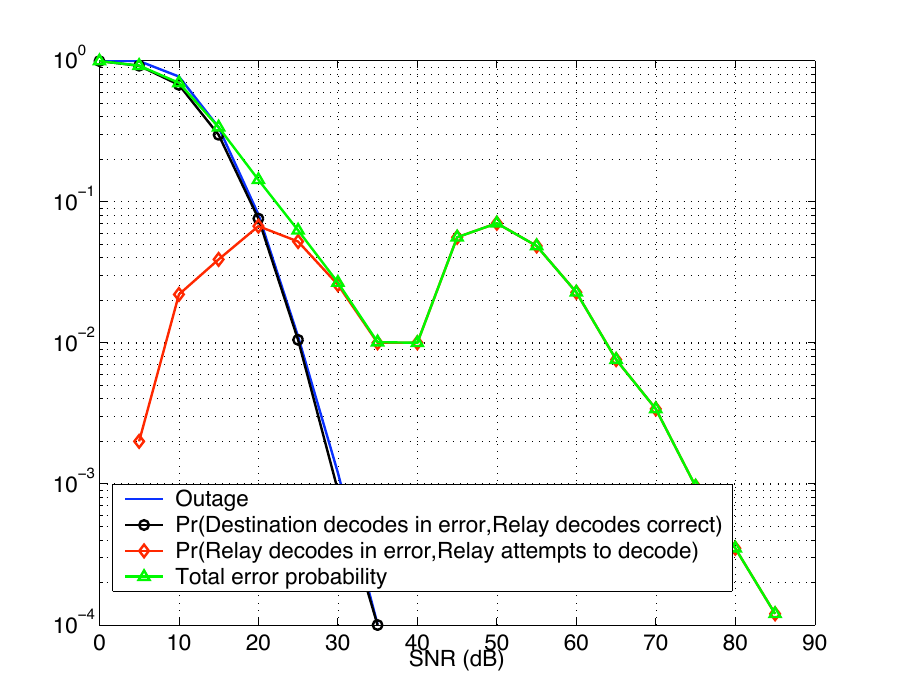}
\caption{$\mathcal{X}_s$ is a rotated QAM code, $T = 1, \ M = 4, \ R
= 4$ bpcu, relay implements $\phi_1(\cdot)$. \label{fig:phi_1}}
\end{center}
\end{figure}

A first choice for $\phi$, which we shall denote $\phi_1$, would be
to allow the relay to decode as soon as the mutual information
between the source and the relay exceeds $MTR$, i.e.,
\[ \phi_1(h) = \min \left\{ M, \left\lceil \frac{MR}{\log (1 + |h|^2 \rho')} \right\rceil \right\}  , \]
where $\rho'$ is the SNR of the source-relay link. This rule is
asymptotically optimal for large $T$,
in fact it coincides with the rule in the original formulation of the DDF protocol
(\ref{ddf-classic}).  For finite $T$, $\phi_1$  is suboptimal since it ignores the actual signal received by the relay, i.e., the
atypical behavior of the noise may dominate the error probability
for short block lengths. As an illustration of the inefficacy of
this decision function at finite block-length, consider the simulation results
in Fig.~\ref{fig:phi_1}. In the simulations to follow, we choose
$\mathcal{X}_s$ to be a rotated QAM code. We will subsequently
compare these results with those obtained by choosing
$\mathcal{X}_s$ to be a permutation code, and observe very similar
trends. We consider ML decoding at both the relay and destination
for all the simulations in this sub-section. Further, we assume that
the source-relay link SNR $\rho'$ is 3 dB above the SNR $\rho$ of
all other links in all our simulations (the X-axis on all our plots
is the SNR $\rho$ in dBs).

The simulations in Fig.~\ref{fig:phi_1} are for the case when
$\mathcal{X}_s$ is a rotated QAM code, $T=1, \ M=4$ and $R = 4$ bits
per channel use (bpcu). The seemingly strange non-monotonic behavior
of the error probability can be understood by the following
intuitive explanation. At low SNRs, the relay hardly ever triggers
before $m = 4$, resulting in $P(\mathcal{E})$ being
dominated by the error probability at the destination
$P(\mathcal{E},\overline{\mathcal{E}}_r)$,
and hence $P(\mathcal{E})$ is large and decreasing.
Then, there is an intermediate region of SNR where the relay attempts to decode, but it
decodes incorrectly with high probability  and causes significant
interference at the destination. Thus $P(\mathcal{E})$ is  dominated by the relay error probability
$P(\mathcal{E}_r)$, and increases in this region. For sufficiently large SNR,
the relay decodes correctly with high probability and therefore helps the destination, thus providing
the required cooperative diversity (slope of the overall error curve at high SNR).
However, this happens at very large gap from the outage probability, that can be regarded as a de-facto optimal performance also
for finite-length codes and not asymptotically high SNR.
This simulation reveals a phenomenon that has been scantily treated in
previous works: the effect of decoding errors at the relay clearly
dominates the overall performance. This fact has often been neglected
since it is neither captured by the $T \rightarrow \infty$ case,
where the atypicality of the noise has no effect and triggering the
relay based on the outage event is exact, nor by the DMT
formulation, that does not capture the gap from outage,
but just the asymptotic error probability curve slope.

\begin{figure}[h]
\begin{center}
\includegraphics[width=15cm]{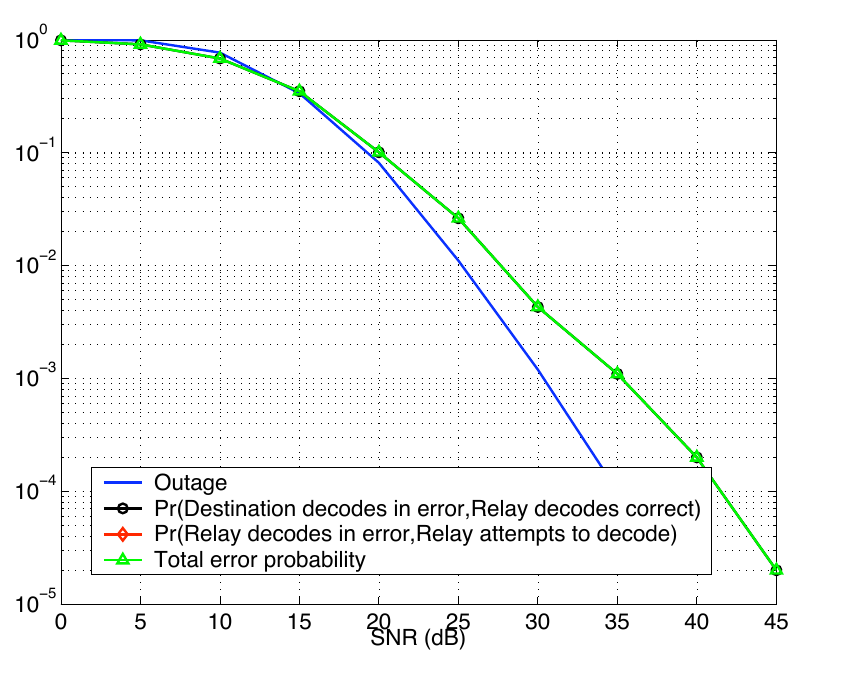}
\caption{$\mathcal{X}_s$ is a rotated QAM code, $T = 1, \ M = 4, \ R
= 4$ bpcu, relay implements $\phi_2(\cdot)$. \label{fig:phi_2}}
\end{center}
\end{figure}

One immediate remedy consists of  adopting a conservative
relay decoding decision function, which we will denote as $\phi_2$, defined as
\[ \phi_2 = \min \left\{ M, \left\lceil \frac{MR}{\log (1 + |h|^2 \rho')} \right\rceil + 1 \right\}. \]
Simulation results using this strategy are shown in
Fig.~\ref{fig:phi_2}, once again for the case of $\mathcal{X}_s$
being a rotated QAM code, $T=1, \ M=4$ and $R = 4$ bpcu.

In this case, the relay error probability is so low that no errors
were recorded in our Monte Carlo simulation (no such curve is shown in
Fig.~\ref{fig:phi_2}). The downside of this strategy however is that
since the relay is over-conservative, it helps the transmitter too late, and
the overall error probability $P(\mathcal{E},\overline{\mathcal{E}}_r)$ suffers from significant
degradation with respect to outage probability.

\begin{figure}[h]
\begin{center}
\includegraphics[width=15cm]{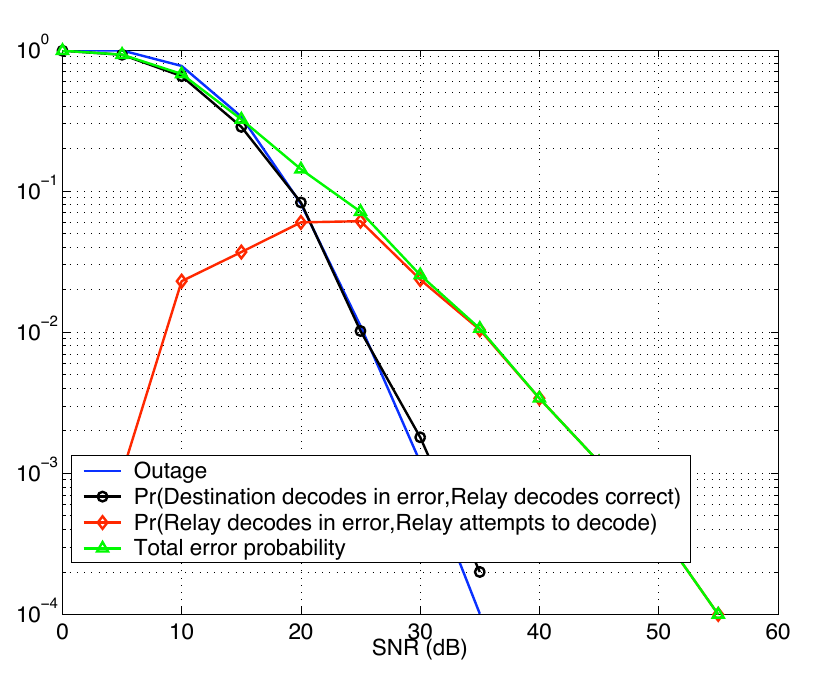}
\caption{$\mathcal{X}_s$ is a rotated QAM code, $T = 1, \ M = 4, \ R
= 4$ bpcu, relay implements $\phi_3(\cdot)$. \label{fig:phi_3}}
\end{center}
\end{figure}

In \cite{MurAzaGam}, the authors prove that there is no loss in DMT
for the DDF protocol using the Alamouti type relay by using the
following relay decoder function $\phi_3$:
\[ \phi_3 = \min \left\{M,  \max \left\{ \frac{M}{2}, \left\lceil \frac{MR}{\log (1 + |h|^2 \rho')} \right\rceil \right\} \right\}, \]
i.e., the relay is allowed to decode and transmit only after the
codeword from the source is at least half-way through. Simulation
results of this protocol shown in Fig.~\ref{fig:phi_3} reveal that
this scheme also suffers from a significant penalty at high SNRs due
to the $P(\mathcal{E}_r)$ term dominating the overall error probability.

In \cite{MurAzaGam}, the authors use an extra layer of
cyclic-redundancy check (CRC) coding to enable the relay to perform
error detection (and wait for another round if incorrect decoding is
detected) - while this strategy is effective in reducing
$P(\mathcal{E}_r)$, there is an inherent loss of rate. In fact, it
is shown in \cite{GamCaiDam_ARQ} for the MIMO-ARQ channel that CRC
is suboptimal in terms of DMT since the undetected error probability
must decrease with SNR at least with the same exponent of error probability itself,
and this requires a number of CRC bits that grow linearly with $\log \text{SNR}$.
The same consideration applies here. Hence, we wish to avoid the use of CRC in order to
detect if the relay decodes in error.

We present a novel strategy to enable error detection at the relay
without further layers of coding at the transmitter. We make use of
a criterion introduced by Forney in \cite{For} in the context of
retransmission (ARQ) protocols to decide whether the decoder is in
error or accept the decoding outcome.
Here, we apply this criterion to the relay decoder, that we refer to as {\em Forney's decision rule}.
Interestingly, Forney's decision rule is similar in essence to the bounded distance decoder that we have considered in
the proof of Theorem \ref{achievability-thm}. However, while the bounded distance decoder is easy to analyze but only
asymptotically optimal, Forney's decision rule
has the remarkable property of striking an optimal balance between the probability of
undetected error at the relay and the probability of rejecting the
decision and waiting for the next slot (probability of decision  ``erasure'', in the language of \cite{For}).
To the best of our knowledge, this decoding decision rule was not proposed before in
the context of relay cooperative communication.

We define the decoding decision function $\phi_F(h,\yv_r)$ using Forney's decision
rule as follows:

\begin{enumerate}
\item If $\phi_1(h) = M$, don't decode and set $\phi_F(h,\yv_r) = M$ (worthless trying to decode if we are in
outage).

\item If $\phi_1(h) = m < M$, decode after the $m^{\text{th}}$ block and apply the
following threshold test. Let $\widehat{\omega}$ denote the outcome of
the relay decoder. Accept the decision and trigger the transmission mode if
\begin{equation} \label{eq:forney_rule}
\frac{p(\yv_{r,0}^{m}| \widehat{\omega}, h)} {\sum_{ \omega \neq
\widehat{\omega}} p(\yv_{r,0}^{m}|\omega, h) } \geq \tau,
\end{equation}
where $\tau$ a suitable threshold set empirically for each SNR.
If the threshold is not exceeded,  wait for the next block and repeat this step
until either the threshold is exceeded or $m = M$.
\end{enumerate}
$\phi_F$ is found to be extremely effective in suppressing the error
probability at the relay without being too conservative and
refraining from helping the destination when possible. Simulation
results for the case when $\mathcal{X}_s$ is a rotated QAM code,
$T=1, \ M=4$ and $R = 4$ bpcu are shown in Fig.~\ref{fig:phi_3}.

\begin{figure}[h]
\begin{center}
\includegraphics[width=15cm]{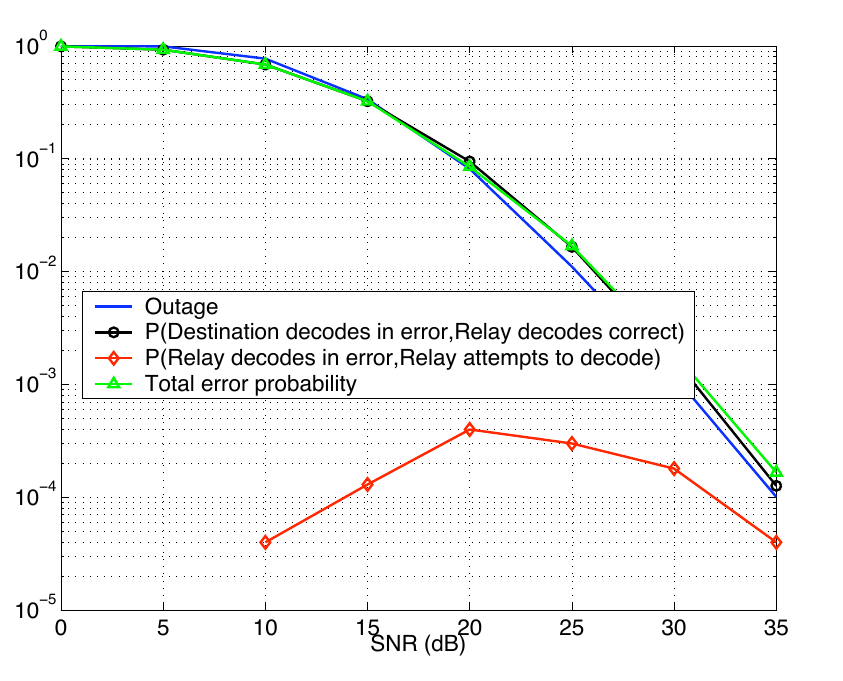}
\caption{$\mathcal{X}_s$ is a rotated QAM code, $T = 1, \ M = 4, \ R
= 4$ bpcu, relay implements $\phi_F(\cdot)$. \label{fig:phi_F}}
\end{center}
\end{figure}
The error probability is within $1$ dB from the corresponding outage
probability.

\begin{figure}[h]
\begin{center}
\includegraphics[width=15cm]{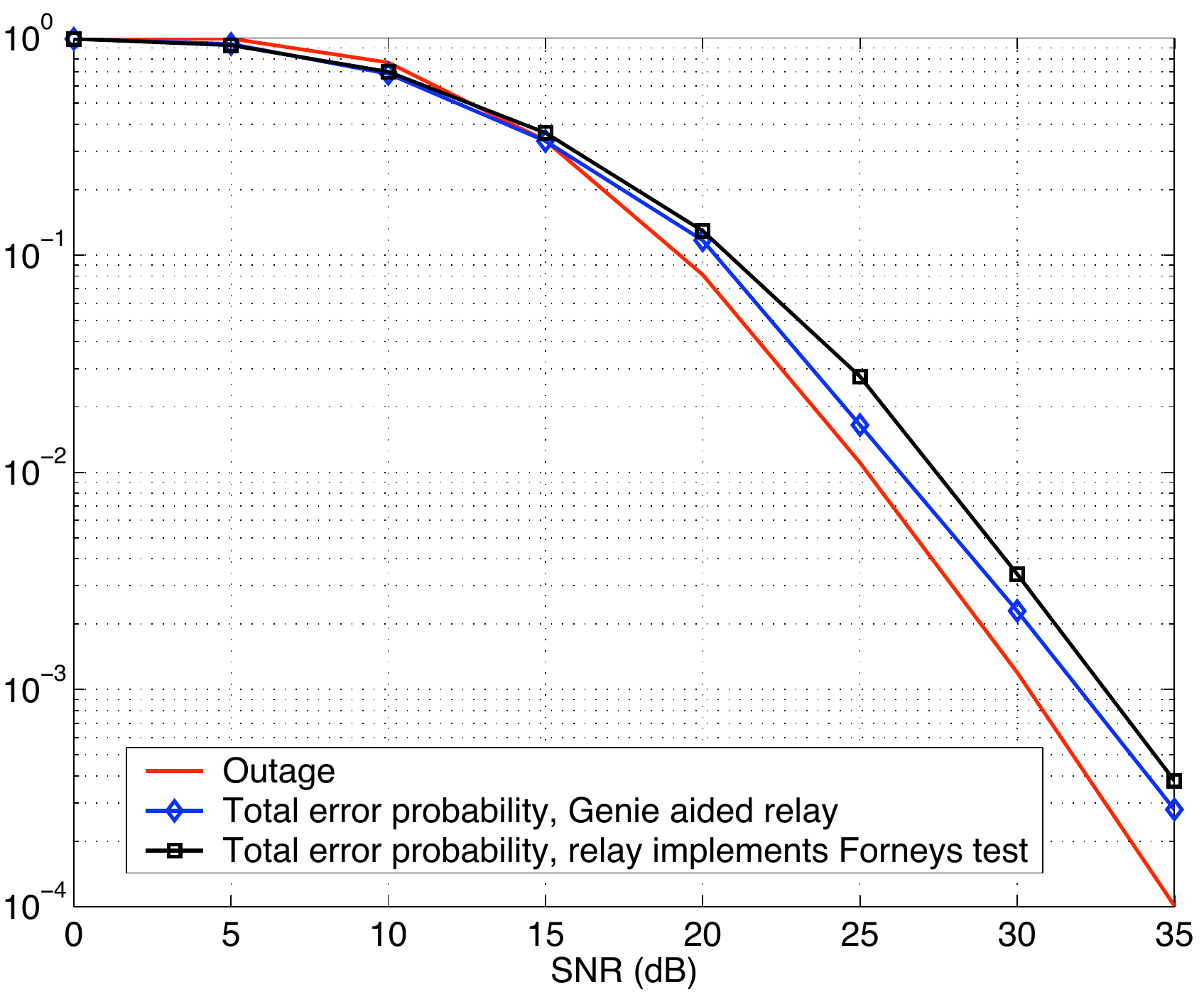}
\caption{$\mathcal{X}_s$ is a permutation code, $T = 1, \ M = 4, \ R
= 4$ bpcu, relay implements $\phi_F(\cdot)$. \label{fig:UDM_phi_F}}
\end{center}
\end{figure}

Fig.~\ref{fig:UDM_phi_F} shows the results when we choose
$\mathcal{X}_s$ to be a permutation code, with $T=1, \ M=4$ and $R =
4$ bpcu.
In this figure we considered two cases: the case of a genie aided relay, where a
genie provides the relay with the source message as soon as the
mutual information exceeds $MTR$, and the case where the relay
performs minimum distance decoding in conjunction with Forney's rule.
The results are similar in flavour to the case where $\mathcal{X}_s$ is a
rotated QAM code, with the permutation code losing $1$ dB with
respect to the rotated QAM code. Notice that despite the DMT optimality, these codes
may perform differently depending on their shaping and coding gain.
In this case, it is apparent that the rotated QAM code outperforms the permutation code,
although they achieve the same diversity.

\subsection{Low complexity MMSE-GDFE Lattice Decoding}

As we saw in \eqref{eq:code_lattice}, the choice of rotated QAM
codes makes $\mathcal{X}_s$ a lattice code. Let $\Lambda$ be the
$2MT$-dimensional lattice corresponding to the generator matrix
$\Gm$ in \eqref{eq:code_lattice}. MMSE-GDFE lattice decoding has
been shown to be DMT optimal for the class of lattice space-time
(LaST) codes over MIMO channels \cite{GamCaiDam_LAST}, and has also
been shown to perform well for deterministic structured LaST
(S-LaST) codes \cite{KumCai_ISIT06,KumCai_ISIT07}. Let
$\mathcal{V}(\Lambda)$ denote the fundamental Voronoi cell of an
$n$-dimensional lattice $\Lambda$ (See \cite{ConSlo} for definitions
relating to lattice theory). The lattice quantization function is
defined by
\[ Q_\Lambda (\yv) \triangleq \arg\min\limits_{\lambdav \in \Lambda} |\yv - \lambdav| \]
and the modulo-lattice function is given by
\[ \yv \mod \Lambda \triangleq \yv - Q_\Lambda (\yv). \]
In the sequel, we will work with the real channel model which is
equivalent to \eqref{eq:recd_relay}, \eqref{eq:recd_destination1}
and \eqref{eq:recd_destination2}, obtained by writing signals
explicitly in terms of their real and imaginary parts (see for
example \cite{Tel} for details regarding the equivalence between
real and complex channel models). By slight abuse of notation, we
will refer to the real equivalent of the complex vectors and
matrices $\xv_s, \yv, \yv_r, \Lambda, \Gm$ using the same notation.
In order to use reduced complexity MMSE-GDFE lattice decoding
\cite{GamCaiDam_LAST}, information needs to be encoded onto cosets
of a sublattice $\Lambda_s$ of $\Lambda$, as follows. Choose
$\Lambda_s = Q \Lambda$, where $Q \in \mathbb{Z}_+$. Thus
$|\Lambda/\Lambda_s| = Q^{2MT}$. Let $\mathcal{C}$ denote the set of
points $\{\Gm \zv \ | \ \zv \in \mathbb{Z}^{2MT}_Q \}$, where $\ZZ_Q
\triangleq \{ 0,1,\hdots,Q-1\}$. The transmitter selects a codeword
$\cv \in \mathcal{C}$, generates a pseudo-random dither signal $\uv$
with uniform distribution over $\mathcal{V}(\Lambda_s)$, and obtains
the transmitted codeword
\[ \xv_s = [\cv - \uv] \mod \Lambda_s. \]
Thus information is encoded onto the cosets of the partition
$\Lambda/\Lambda_s$: $\xv_s$ is a coset representative of the coset
onto which the information is encoded, and belongs to the
fundamental Voronoi region of $\Lambda_s$. Let $\mathcal{C}_\omega$
denote the coset of $\Lambda_s$ in $\Lambda$ onto which the
information corresponding to message $\omega$ is encoded. From
\eqref{eq:recd_relay}, \eqref{eq:recd_destination1} and
\eqref{eq:recd_destination2} the (real equivalent) received signals
$\yv_r = \yv_{r,0}^{m}$ for $\Mc = m$ at the relay and $\yv_s =
\yv_{s,0}^{M}$ at the destination may be written as
\begin{equation} \label{eq:recd_relay_vec}
\yv_r = \Hm_r \xv_s + \vv
\end{equation}
and
\begin{equation} \label{eq:recd_dest_vec}
\yv_s = \Hm \xv_s + \wv,
\end{equation}
where $\Hm_r \in \mathbb{C}^{2 m T \times 2MT}$ and $\Hm \in
\mathbb{C}^{2MT \times 2MT}$ denote the (real) equivalent channels
at the relay and destination, and $\vv$ and $\wv$ denote the (real
equivalent) noise at the relay and destination respectively. Notice
from \eqref{eq:recd_relay_vec} that decoding at the relay
corresponds to solving an under-determined system of linear
equations. We follow the approach of \cite{DamGamCai_underdet_CISS}
in this case, where it was shown how MMSE-GDFE lattice decoding may
be used to efficiently solve under-determined systems of linear
equations. We focus on decoding at the relay in the sequel, the
decoder at the destination is identical upon replacing the relevant
signals and parameters at the relay with those at the destination.
Let $\Fm$ and $\Bm$ denote the forward and backward filters of the
MMSE-GDFE (see for example \cite{GamCaiDam_LAST} for the definition
of these matrices in terms of $\Hm_r$ and the relay SNR). The relay
produces the modified observation
\[ \yv_r'  \triangleq \Fm \yv_r + \Bm \uv,\]
and computes
\[ \widehat{\zv} = \arg \min\limits_{\zv \in \mathbb{Z}^{2MT}} |\yv_r' - \Bm \Gm \zv|^2. \]
The relay then decides in favor of the coset
$\mathcal{C}_{\widehat{\omega}}$ that contains the point
\[ [\Gm \widehat{\zv}] \mod \Lambda_s. \]
In order to work with the lattice coding and decoding scheme,
Forney's decision rule  \eqref{eq:forney_rule} needs to be modified
to take into account the fact that information is encoded onto
cosets as against points in the lattice.   Encoding information into
cosets is equivalent to consider a modulo-$\Lambda_s$ channel with
output $\yv_r'$ modulo $\Bm\Lambda_s$. Hence, the relevant
likelihood function is given by
\[ \widetilde{p}(\yv_r'|\omega, h) = \sum_{\lambda_s \in \Lambda_s}
p_{\widetilde{w}}(\yv_r' - \Bm (\cv_{\omega} + \lambda_s) | h) \]
with domain $\yv_r' \in \Vc(\Bm \Lambda_s)$, where $\cv_\omega$ is a
coset representative of $\Cc_\omega$ and where $p_{\widetilde{w}}(
\wv|h)$ denotes the pdf of the noise induced by the
modulo-$\Lambda_s$ channel with the dithering, that is,
$\widetilde{\wv} = \yv_r' - \Bm \xv_s$ where $\xv_s$ is the
transmitted signal. Unfortunately, $p_{\widetilde{w}}$ is difficult
if not impossible to determine in closed form. However, a good
practical choice that works well for good shaping lattices is to let
$p_{\widetilde{w}}$ be a Gaussian pdf with i.i.d. components $\sim
\Nc(0, \sigma_v^2/2)$ (see \cite{GamCaiDam_LAST} for a theoretical
asymptotic justification of Gaussianity in this context). Then, the
proposed modification of Forney's decision rule
\eqref{eq:forney_rule} for the lattice MMSE-GDFE decoder is given
by: accept $\widehat{\omega}$ at time $m$ if
\begin{equation} \label{eq:forney_rule_modified}
\frac{\sum_{\lambda_s \in \Lambda_s} p_{\widetilde{w}} (\yv'_{r} -
\Bm (\cv_{\widehat{\omega}} + \lambda_s) | h)} {\sum_{\omega \neq
\widehat{\omega}} \sum_{\lambda_s \in \Lambda_s} p_{\widetilde{w}}
(\yv'_{r} - \Bm ( \cv_{\omega} + \lambda_s) | h) } \geq \tau,
\end{equation}
where, again $\tau$ is a suitable threshold set empirically for each
SNR. The infinite sums at numeration and denominator can be safely
truncated by restricting to a number of most likely lattice points,
which may be done as follows. Generate a list of $\Nc = \{ \Bm \lambda_i : \lambda_i \in \Lambda \}_{i=1}^{\Nc}$
of lattice points of the lattice generated by $\Bm \Gm$
that are closest to $\yv_r'$. Such a list may be generated, for example, by using a standard lattice decoder
with a sufficiently large search radius.
For any given message $\omega$, check whether $\lambda_i$ belongs to the coset $\cv_{\widetilde{\omega}} + \Lambda_s$.
If yes, then this point makes a contribution towards the numerator of
\eqref{eq:forney_rule_modified}, else towards the denominator.
If there exists $\widehat{\omega}$ for which the corresponding ratio crosses the threshold $\tau$ then accept the decision, otherwise
reject and wait for the next slot.

\begin{figure}[h]
\begin{center}
\includegraphics[width=15cm]{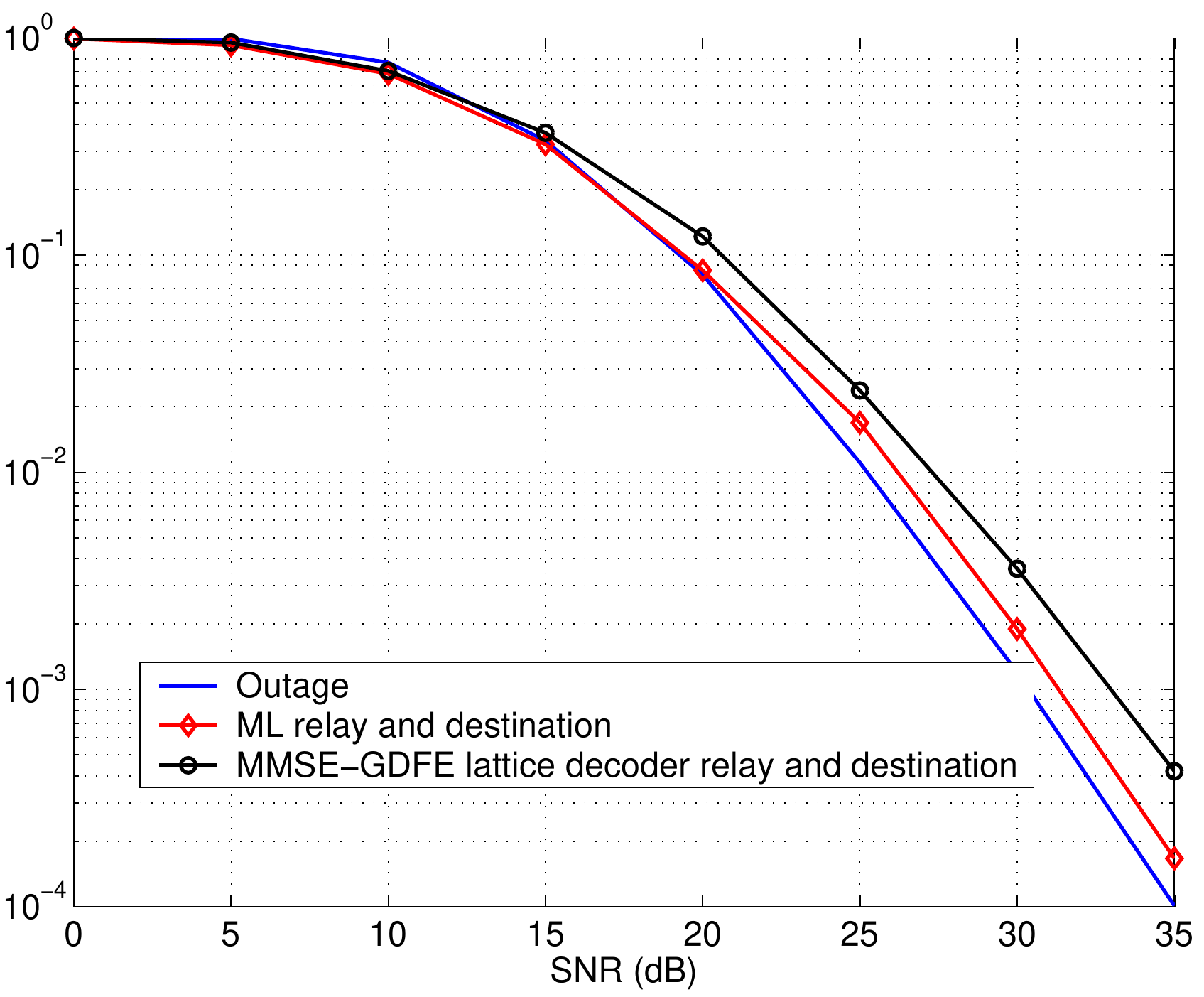}
\caption{$\mathcal{X}_s$ is a rotated QAM code, $T = 1, \ M = 4, \ R
= 4$ bpcu, relay implements Forney's or modified Forney's rule.
\label{fig:MMSE_forney}}
\end{center}
\end{figure}

The modified Forney's rule in \eqref{eq:forney_rule_modified} is
seen to be quite effective for the case when MMSE-GDFE lattice
decoding is performed at both the relay and the destination. The
simulations in Fig.~\ref{fig:MMSE_forney} compares the performance
of rotated QAM codes with $T = 1, \ M = 4, \ R = 4$ bpcu that use
Forney's rule \eqref{eq:forney_rule} with ML decoding and modified
Forney's rule \eqref{eq:forney_rule_modified} with MMSE-GDFE lattice
decoding. The low complexity lattice decoder tracks the ML performance within $1$ dB.

\section{Conclusion}

We presented a characterization of the achievable DMT of the
single-relay DDF protocol for finite block length. Our achievability
proof yields insight on the design of actual coding schemes. In
particular, we stressed the importance of a relay decoding function
that check the reliability of its decision, in order not to jam the
destination. Also, we showed that the destination need not be aware
of the relay decision time, since a GLRT-based decoder achieves
optimal DMT performance. This may have some impact on the design of
practical DDF protocols, since it essentially shows that no
complicated side-information channel needs to be implemented in
order to explicitly notify the destination about when the relay
starts transmitting.

In our proofs, we considered a bounded distance decoder at the relay
and an ensemble of random Gaussian codes. Then, we constructed
practical and very simple codes based on lattices (rotated QAM
constellations) and permutation codes. We demonstrated via
simulation that the impact of undetected decoding errors at the
relay may be huge. In order to tackle this problem, we have proposed
the use of Forney's decision rejection rule, that proves to be very
effective.

Finally, we have investigated the use of a reduced complexity MMSE-GDFE lattice decoder and modulo-$\Lambda$
lattice codes, that yields the well-known low-complexity decoding even at the relay.
It should be remarked that the relay invariably has to decode an undetermined linear system, therefore
standard sphere decoding algorithms fail.

A few comments relating to future work are in order here. We have
not exploited the low-complexity quantization-based decoding
approach for permutation codes \cite{TavVis_IT}, owing to the fact
that it is not completely clear as to how we can apply a good
decision rejection rule at the relay in this case. Another
interesting problem relates to code design for finite but large $T$;
in this case, neither the constructions presented in this paper, nor
those in \cite{EliKum,EliKum_arxiv} are fully controllable in terms
of coding gain and both entail very high decoding complexity.
Concatenation of short codes based on rotated QAM constellations or
permutation codes with some form of outer coding (along the lines of
\cite{FabCai}) may prove to be appropriate for this scenario.

\appendices

\section{Separated Relay Activity Detection \label{app:RAD}}

In this Appendix we treat a side problem. An intuitive
low-complexity scheme for detecting the relay decision time consists
of treating $\Mc$ as a random parameter, and use ML detection by
disregarding the structure of the channel codes. Intuitively, the
destination should be able to detect a transition in the received
power, between the listening phase and the transmission phase of the
relay. This approach is referred to as separated {\em Relay Activity
Detection} (RAD), since the decision time and the source codeword
are separately decoded, in contrast with the GLRT decoder analyzed
in the proof of Theorem~\ref{achievability-thm}. We shall show that
separated RAD yields no performance loss when we consider the limit
of $T \rightarrow \infty$.  On the contrary, it is suboptimal and
actually may perform very poorly when limits are taken in the
reverse order, that is, for each finite $T$ we consider the
performance as SNR gets large.

We assume that the source uses an i.i.d. random Gaussian code and
the relay implements the Alamouti-DDF scheme. As before, let $\Mc$
denote the decision time. An ML decision time detector that is
ignorant of the codebooks treats the channel input as a random
Gaussian signal. The detection rule is given by
\[ \widehat{\Mc} = \arg \max\limits_{m} p (\yv | \Mc = m, g_1,g_2).\]
where $p (\yv | \Mc = m, g_1,g_2)$ shall be denoted in the following
simply by $p (\yv | m)$ for simplicity, and it is given by
\begin{eqnarray} \label{eq:naive_prob_distrib}
p(\yv|m) & = & \frac{1}{\left[ \pi (|g_1|^2 \rho + 1 ) \right]^{mT}}
\exp \left( - \frac{|\yv_0^{m}|^2}{|g_1|^2 \rho + 1} \right)\\
& & \cdot \frac{1}{ \left\{ \pi \left[ (|g_1|^2 + |g_2|^2)\rho + 1
\right] \right\}^{(M-m)T} }
\exp \left( \frac{-|\yv_{m}^{M}|^2}{(|g_1|^2+|g_2|^2) \rho + 1} \right).
\end{eqnarray}
Suppose $\Mc = m$, we define the pairwise error event
\[ \{ m \rightarrow m' \} \triangleq \left\{  \frac{p(\yv | m')}{p (\yv | m)}
\geq 1 \right \}. \] The detector error probability is lower bounded
by
\[ P(\Mc \neq \widehat{\Mc}) \geq \max_{m\neq m'} P( m \rightarrow m'), \]
and is upper bounded by the union bound
\[ P(\Mc \neq \widehat{\Mc}) \leq (M-1) \max_{m \neq m'}
P( m \rightarrow m'). \]
Hence, we shall study the diversity exponent of $P( m \rightarrow
m')$ for general $m \neq m'$. If this does not depend on $m, m'$ we
have determined the diversity exponent of the separated RAD.

\subsection{Infinite block-length}

If $\Mc = m$ and $T \rightarrow \infty$, the law of large numbers
yields the almost sure convergence of the limits:
\begin{eqnarray*}
\frac{1}{T} |\yv_{n-1}^{n}|^2 & \rightarrow&  |g_1|^2 \rho + 1, \;\;
1 \leq n \leq m
\end{eqnarray*}
and
\begin{eqnarray*}
\frac{1}{T} |\yv_{n-1}^{n}|^2 \ \rightarrow \ (|g_1|^2 + |g_2|^2)
\rho + 1, \;\; m+1 \leq n \leq M.
\end{eqnarray*}
Thus, for large $T$ we have
\[
p (\yv | m) \approx \exp \left\{ -MT -mT \log \left[ \pi (|g_1|^2
\rho + 1) \right]  - (M-m)T \log \left[ \pi ((|g_1|^2
+ |g_2|^2)\rho+1) \right] \right\}.
\]
Consider the case $m' > m$ (the other case follows in the same way
and it is omitted for brevity). We have
\begin{eqnarray*}
p(\yv|m') & \approx & \exp \left\{ - m T - (M-m')T  -(m'-m)T \frac{(|g_1|^2+|g_2|^2)\rho+1}{|g_1|^2 \rho+1} - \right . \\
& & \left . - m'T \log \left[ \pi (|g_1|^2 \rho+1) \right] - (M-m')T \log \left[
\pi\left( (|g_1|^2+|g_2|^2)\rho+1 \right) \right] \right\}.
\end{eqnarray*}
After some simplifications, the pairwise error probability for $T
\rightarrow \infty$ is given by
\begin{equation}
P (m \rightarrow m'| g_1,g_2) = P \left ( 1 - X_1 + \log X_1
\geq 0 \right ),
\end{equation}
where we let
\[ X_1 = \frac{(|g_1|^2+|g_2|^2)\rho+1}{|g_1|^2 \rho+1}. \]
Since $\log x \leq x - 1 \ \forall \ x \geq 0$, we see that $\{m
\rightarrow m'\}$ can occur only if $|g_2|^2 = 0$, which is an event
of measure $0$. Therefore, we conclude that $P(m \rightarrow m')
\downarrow 0$ for any fixed $\rho$, as $T \rightarrow \infty$. This shows
that for the infinite $T$ case, even a very simple separated RAD
scheme at the destination yields perfect knowledge of the relay
decision time without any need of a side information channel that
involves some protocol overhead.

\subsection{Finite block-length}

We now fix $T$ to be an arbitrary finite value and study the
diversity exponent of $P(m \rightarrow m')$ as $\rho \rightarrow
\infty$. Again, we consider only the case $m' > m$. The likelihood
function for the hypothesis $m'$ when $\Mc = m$ is given by
\begin{eqnarray*}
p(\yv|m') & = & \exp \left( -m' T \log \left[ \pi (|g_1|^2 \rho + 1 )
\right]   - (M-m')T \log \pi \left[ (|g_1|^2 + |g_2|^2)\rho + 1 \right] \right . \\
& & \left. - \frac{|\yv_0^{m}|^2 + |\yv_{m}^{m'}|^2}{|g_1|^2 \rho +
1} - \frac{|\yv_{m'}^{M}|^2}{ (|g_1|^2+|g_2|^2) \rho + 1} \right).
\end{eqnarray*}
After some algebra, we find that
\begin{equation}
\label{pep} P(m \rightarrow m'| g_1,g_2) = P \left ( \chi \leq
\frac{(m'-m)T}{X_2} \log(1 + X_2) \right ),
\end{equation}
where
\[ \chi = \frac{\left |\yv_{m}^{m'} \right |^2}{1 + (|g_1|^2 + |g_2|^2)\rho} \]
is a central chi-squared random variable with $2T(m'-m)$ degrees of
freedom and mean $T(m'-m)$, and we define
\[ X_2 = \frac{|g_2|^2\rho}{|g_1|^2\rho+1}.\]
As an aside, notice that $\frac{1}{x} \log(1 + x)$ is a decreasing
function of $x$ that is less than $1$ for all $x > 0$, and
approaches 1 for $x \downarrow 0$. Therefore, the term
$\frac{(m'-m)T}{X_2} \log(1 + X_2)$ in (\ref{pep}) is always
strictly less than $\EE[\chi] = (m'-m)T$  for all $|g_2| > 0$.
Therefore, as an application of the the large deviation theorem
\cite{DemZei}, we find that $P(m \rightarrow m') \downarrow 0$
exponentially with $T$ for all finite $\rho$ and $|g_2|>0$. Thus, we
recover in a more rigorous way the result obtained before by letting
$T \rightarrow \infty$ directly in the detector decision metric.

Returning to the case of finite $T$, we have using well-known
properties of the chi-squared distribution that
\[ P(\Xc \leq u) = \frac{1}{((m'-m)T)!} u^{(m'-m)T} + O(u^{(m'-m)T + 1}) \]
for small $u$, and obviously
\[ P(\Xc \leq u) = O(1) \]
when $u = \beta (m' -m)T$ for some constant $\beta > 0$. Fix an
arbitrary $0 < \beta < 1$. From what was said before, there exists
an $x_2 > 0$ such that $\frac{1}{x_2} \log(1 + x_2) = \beta$. Hence,
consider the event
\begin{eqnarray} \label{Edef}
\Ec(\rho,\beta) & = & \left \{ X_2 \leq x_2 \right \}   \nonumber \\
& = & \left \{ |g_2|^2 \rho \leq x_2 (1 + |g_1|^2 \rho) \right \}.
\end{eqnarray}
It is clear that for all $(g_1,g_2) \in \Ec(\rho,\beta)$, the
pairwise error probability $P(m \rightarrow m'| g_1,g_2)$ in
(\ref{pep}) is exponentially equivalent to a constant as $\rho
\rightarrow \infty$, i.e.,
\[ P(m \rightarrow m'| g_1,g_2) \; \doteq \; \rho^0, \;\;\; (g_1,g_2) \in \Ec(\rho,\beta). \]
Averaging with respect to $g_1,g_2$, and using the standard variable
substitution $|g_1|^2 = \rho^{-\alpha_1}$, $|g_2|^2 =
\rho^{-\alpha_2}$, we find
\begin{eqnarray*}
P(m \rightarrow m') &\dot\geq& \int_{\Ec} \rho^0 e^{
-\rho^{-\alpha_1} -\rho^{-\alpha_2} } \rho^{-\alpha_1-\alpha_2}
d\alpha_1 d\alpha_2\\
&\doteq& \int_{\Ec'} \rho^{-\alpha_1-\alpha_2} d\alpha_1 d\alpha_2,
\end{eqnarray*}
where, from (\ref{Edef}),
\[ \Ec' = \left \{ (\alpha_1,\alpha_2) \in \RR_+^2 : 1 - \alpha_2 \leq [1 - \alpha_1]_+ \right \}. \]
Using Varadhan's lemma, we find that the diversity exponent of the
pairwise error probability is given by
\[  \Delta = \inf_{(\alpha_1,\alpha_2) \in \Ec'} \left \{ \alpha_1 + \alpha_2 \right \}  = 0, \]
since the point $\alpha_1 = 0,\alpha_2 = 0$ belongs to the boundary
of the region $\Ec'$.

This shows that for any finite $T$, a separated RAD scheme based on
optimal (Maximum Likelihood) detection of the relay decision time
$\Mc$ that ignores the codebook structure and treats the transmitted
signals as random processes is very suboptimal. In fact, the
probability of error of such a scheme is constant with SNR and
eventually will dominate the performance of the whole destination
decoder.

In some way, this result shows that the joint detection of the relay
decision time and of the information message is {\em necessary} in
order to achieve the optimal (infinite $T$) DDF DMT.

\end{document}

%% file: macros.tex
\newfont{\bbb}{msbm10 scaled 500}

\newfont{\bb}{msbm10 scaled 1100}
\newcommand{\CC}{\mbox{\bb C}}
\newcommand{\RR}{\mbox{\bb R}}

\newcommand{\ZZ}{\mbox{\bb Z}}
\newcommand{\FF}{\mbox{\bb F}}

\newcommand{\EE}{\mbox{\bb E}}

\newcommand{\av}{{\bf a}}
\newcommand{\bv}{{\bf b}}
\newcommand{\cv}{{\bf c}}

\newcommand{\uv}{{\bf u}}
\newcommand{\wv}{{\bf w}}
\newcommand{\vv}{{\bf v}}
\newcommand{\xv}{{\bf x}}
\newcommand{\yv}{{\bf y}}
\newcommand{\zv}{{\bf z}}

\newcommand{\Am}{{\bf A}}
\newcommand{\Bm}{{\bf B}}

\newcommand{\Fm}{{\bf F}}
\newcommand{\Gm}{{\bf G}}
\newcommand{\Hm}{{\bf H}}
\newcommand{\Id}{{\bf I}}

\newcommand{\Mm}{{\bf M}}

\newcommand{\Rm}{{\bf R}}

\newcommand{\Ac}{{\cal A}}
\newcommand{\Bc}{{\cal B}}
\newcommand{\Cc}{{\cal C}}
\newcommand{\Dc}{{\cal D}}
\newcommand{\Ec}{{\cal E}}

\newcommand{\Mc}{{\cal M}}
\newcommand{\Nc}{{\cal N}}
\newcommand{\Oc}{{\cal O}}

\newcommand{\Rc}{{\cal R}}
\newcommand{\Sc}{{\cal S}}

\newcommand{\Uc}{{\cal U}}

\newcommand{\Vc}{{\cal V}}
\newcommand{\Xc}{{\cal X}}

\newcommand{\lambdav}{\text{\boldmath$\lambda$}}

\newcommand{\xiv}{\hbox{\boldmath$\xi$}}

\renewcommand{\det}{{\hbox{det}}}
\newcommand{\trace}{{\hbox{tr}}}

\renewcommand{\arg}{{\hbox{arg}}}

\renewcommand{\Re}{{\rm Re}}